\documentclass[pra,aps,superscriptaddress,footinbib,twocolumn]{revtex4-1}
\usepackage{mathrsfs}
\usepackage{amsfonts}
\usepackage{amssymb}
\usepackage{graphicx}
\usepackage{subfigure}
\usepackage[usenames,dvipsnames]{color}
\usepackage[colorlinks=true,citecolor=blue,linkcolor=magenta]{hyperref}
\usepackage{lmodern}
\usepackage{dsfont}
\usepackage{natbib}
\usepackage{bm}
\usepackage{pgf}
\usepackage{calligra}

\usepackage{algorithmicx}

\usepackage{amsmath} 

\usepackage{algorithm}
\usepackage{algpseudocode}
\newtheorem{theo}{Theorem}
\newtheorem{lemma}{Lemma}
\newcommand{\Tr}{\mathrm{Tr}}
\newcommand{\1}{{j,k_j}}
\newcommand{\2}{{i,k_i}}
\newcommand{\3}{{l,k_l}}
\newcommand{\4}{{m,k_m}}

\newcommand{\ket}[1]{\vert{ #1 }\rangle}

\begin{document}
\title{Shadow-based quantum subspace algorithm for the nuclear shell model}

\author{Ruyu Yang}
\affiliation{Graduate School of China Academy of Engineering Physics, Beijing 100193, China}

\author{Tianren Wang}
\affiliation{Graduate School of China Academy of Engineering Physics, Beijing 100193, China}


\author{Bing-Nan Lu}
\affiliation{Graduate School of China Academy of Engineering Physics, Beijing 100193, China}

\author{Ying Li}
\affiliation{Graduate School of China Academy of Engineering Physics, Beijing 100193, China}

\author{Xiaosi Xu}
\email{xsxu@gscaep.ac.cn}
\affiliation{Graduate School of China Academy of Engineering Physics, Beijing 100193, China}

\begin{abstract}

In recent years, researchers have been exploring the applications of noisy intermediate-scale quantum (NISQ) computation in various fields. One important area in which quantum computation can outperform classical computers is the ground state problem of a many-body system, e.g., the nucleus.  
However, using a quantum computer in the NISQ era to solve a meaningful-scale system remains a challenge. 
 To calculate the ground energy of nuclear systems, we propose a new algorithm that combines classical shadow and subspace diagonalization techniques. Our subspace is composed of matrices, with the basis of the subspace being the classical shadow of the quantum state.
We test our algorithm on nuclei described by Cohen-Kurath shell model and USD shell model. We find that the accuracy of the results improves as the number of shots increases, following the Heisenberg scaling.

\end{abstract}
\maketitle

\section{Introduction}
Nuclear \textit{ab initio} calculation, e.g., solving the nuclear ground state from the bare nucleon-nucleon interactions, is notoriously hard on classical computers due to their exponential complexity. On the other hand, quantum computing has emerged as a promising new paradigm that is expected to provide more effective solutions for these problems~\cite{stevensoncomments}. Nonetheless, noise is a significant issue in quantum computing due to hardware limitations~\cite{preskill2018quantum}.
 To make better use of existing Noisy Intermediate-Scale Quantum (NISQ) devices, several classical-quantum hybrid algorithms have been proposed~\cite{endo2021hybrid,li2017efficient,bharti2021noisy,tilly2022variational,tang2021qubit,wang2019accelerated}. Among those, the variational quantum eigensolver (VQE) is a major class of algorithms~\cite{endo2021hybrid,bharti2021noisy,tilly2022variational,tang2021qubit,wang2019accelerated,rattew2019domain,peruzzo2014variational}. Recently, VQE has been applied to solve static problems in nuclear systems~\cite{romero2022solving,dumitrescu2018cloud,stevensoncomments,di2021improving,siwach2021quantum,yeter2020practical,lv2022qcsh}. 
 The ground state of the deuteron was first solved on a cloud quantum computing platform using VQE~\cite{dumitrescu2018cloud}.
Since then, researchers have applied this algorithm to the Lipkin model and shell model (SM) as well~\cite{romero2022solving,stetcu2021variational,kiss2022quantum,cervia2021lipkin}. Along with the VQE algorithm, the imaginary time evolution was also employed in quantum computing to calculate the nuclear ground state energy~\cite{yeter2020practical,lv2022qcsh}. This method involves projecting an initial trial state onto the ground state using an imaginary time evolution operator. Another approach for solving the ground state problem is based on quantum subspace diagonalization (QSD)~\cite{cortes2022quantum,cohn2021quantum,stair2020multireference,klymko2022real,shen2022real,yeter2020practical,huggins2020non,seki2021quantum,kirby2022exact,tkachenko2022quantum,francis2022subspace,yoshioka2022generalized}. This method involves selecting a subspace of the Hilbert space that consists of wavefunctions, and then diagonalizing the Hamiltonian within that subspace. The choice of wavefunctions varies, and there are different ways to generate them. One approach is to start from an initial state that has a finite overlap with the ground state and then evolve the real-time Hamiltonian to generate additional wavefunctions~\cite{klymko2022real}. By selecting different evolution times, a set of wavefunctions can be constructed to form a subspace. Another approach is to use imaginary time evolution to generate a subspace. In~\cite{yeter2020practical}, the Lanczos algorithm was utilized to compute the ground state energy of the deuteron by employing such a subspace method. The definition of the subspace is not limited to wavefunctions, as shown in ~\cite{yoshioka2022generalized}, where the subspace is defined as a polynomial of the density matrix, enabling error mitigation.

Previous algorithms have encountered difficulties in achieving highly accurate results, as the optimization of VQE is a known NP-hard problem~\cite{bittel2021training}, and can become trapped in local minima. Furthermore, the problem of vanishing gradients can arise when the number of qubits is large~\cite{holmes2022connecting}. In addition, accurate imaginary-time evolution requires a deeper and more complex circuit than that required for real-time evolution. When using the QSD algorithm, it is necessary to truncate the overlap matrix S of the subspace to remove singular values below a certain threshold~\cite{epperly2022theory}, and failure to do so will result in a high number of shots being required to reduce the variance of the results; however, truncation itself introduces an error. In general, imaginary-time evolution and QSD also require the use of Hadamard tests to measure the cross-terms $\langle \psi_i|H| \psi_j\rangle$ and the overlap $\langle\psi_i|\psi_j\rangle$, further complicating the circuit. Note that in cases where the particle number of the system is conserved, it may not be necessary to use the Hadamard test. However, even in such cases, it still requires a deeper circuit than simply preparing $|\psi_i\rangle$ \footnote{When measuring the overlap $\langle\psi_i|\psi_j\rangle$, where $|\psi_i\rangle$ and $|\psi_j\rangle$ can be prepared from the initial state $|0\rangle$ and unitary circuits $U_i$ and $U_j$, both $U_i^{\dagger}$ and $U_j$ must be constructed in the circuit~\cite{lu2021algorithms,o2021error}.}.

To overcome these challenges, we introduce a novel class of subspaces. Our approach involves using real-time evolution to generate a set of quantum states, and then constructing their classical shadows. Using these classical shadows, we can construct a subspace composed of matrices, and diagonalize the Hamiltonian within this subspace. Unlike VQE, our algorithm does not require an optimization process but only needs to prepare a state with a finite overlap with the ground state. Furthermore, our method is less susceptible to statistical fluctuations compared to QSD, which eliminates the need to truncate $S$. Additionally, we observe that the measurement accuracy increases with shots and approaches the Heisenberg limit.

To evaluate the effectiveness of our algorithm, we have calculated the ground states of various nuclear systems, including ${}^6\mathrm{He},{}^6\mathrm{Li},{}^7\mathrm{Bo},{}^8\mathrm{Li},{}^{18}\mathrm{O},$ and ${}^{18}\mathrm{F}$ using SM. We have investigated the impact of the subspace size and number of shots on the accuracy of our method through numerical analysis.

Our paper is structured as follows: In Section~\ref{sec:Re}, we provide an overview of related work related to nuclear systems. In Section~\ref{sec:Nu}, we introduce the nuclear SM, which may be unfamiliar to quantum computing researchers. In Section~\ref{sec:Cl}, we review the classical shadow. In Section~\ref{sec:Mo}, we present the modified QSD algorithm and discuss its working principles. In Section~\ref{sec:Fu}, we present a comprehensive algorithmic framework for our approach. In Section~\ref{sec:Si}, we report the results of our numerical experiments. Finally, we conclude our work in Section~\ref{sec:Co}.



\section{Related works}
\label{sec:Re}

VQE is one type  of variational algorithm that is used to find the ground state of a given Hamiltonian. 
In VQE, a parameterized quantum circuit is constructed. During the training process, these variables are iteratively updated to minimize the expectation value of the Hamiltonian.
\begin{equation}
    \langle H(\theta)\rangle = {\langle 0 |U(\theta) HU^{\dagger}(\theta)| 0\rangle}.
\end{equation}

VQE was the first NISQ algorithm used for calculating the binding energy of nuclear systems~\cite{dumitrescu2018cloud}. This method was initially used for the simplest nuclear model and has been extended to more sophisticated models such as the Lipkin and nuclear SM~\cite{romero2022solving, stetcu2021variational, kiss2022quantum, cervia2021lipkin}. In particular, the authors of Ref.~\cite{romero2022solving} utilized an adaptive strategy to generate a suitable ansatz circuit. One potential drawback of VQE is the possibility of getting stuck in a local minimum during parameter updates.  


In addition to variational algorithms, the ground state energy of nuclear systems was also solved by imaginary time evolution algorithms~\cite{yeter2020practical}. This algorithm can not only find the ground state but also prepare the thermal state. Another related work uses the method called full quantum eigensolver~\cite{lv2022qcsh}, which calculates the ground states of various nuclei using the harmonic oscillator basis. The full quantum eigensolver can be considered as a first-order approximation of imaginary time evolution. 

A state that has 
undergone imaginary time evolution becomes
\begin{equation}
    |\Psi(\beta)\rangle=\frac{\mathrm{e}^{-\beta H}|\Psi(0)\rangle}{\| \mathrm{e}^{-\beta H}|\Psi(0)\rangle \|},
    \label{eq:imaginary}
\end{equation}
where $\| \mathrm{e}^{-\beta H}|\Psi(0)\rangle \|$ denotes the normalization factor.
If there is a finite overlap between $\ket{\Psi(0)}$ and the ground state, Eq.~\ref{eq:imaginary} will exponentially approach the ground state as $\beta$ increases. Unfortunately, it's usually difficult to implement the non-unitary operator $\mathrm{e}^{-\beta H}$ with quantum circuits. One approach is to achieve such evolution through variational algorithms~\cite{mcardle2019variational}. The authors variationally prepared the states after a short-time imaginary time evolution. 
 Apart from this, there are two main approaches to implementing this operator. One method is to use a unitary evolution to approximate this non-unitary evolution~\cite{motta2020determining}. While the approach used state tomography to find a suitable unitary evolution, the computational cost grows exponentially with the system's correlation length. Another approach is to use a linear combination of unitary operators to approximate the non-unitary operator~\cite{terashima2005nonunitary, wan2022randomized}.

\section{The nuclear shell model}
\label{sec:Nu}
Nuclear force can be derived from the fundamental theory of QCD. However, in practical calculations, these forces are usually too complex and inaccurate. 
As alternatives, one can work in a limited Hilbert space and write down a phenomenological interaction intended to reproduce the experimental data. 
One example is the nuclear SM. The SM usually involves an inert core consisting of closed shells, while the remaining valence nucleons fill the orbitals in the open shells. 
In SM a single-particle state is characterized by four good quantum numbers: angular momentum $J$ and its 3rd component $J_z$, isospin $T$ and its 3rd component $T_z$.

The Hamiltonian in the SM can be represented in a second quantization form that includes kinetic energy and two-body interactions. We use lowercase letters $a$, $b$, $c$, and $d$ to denote spherical orbitals. Taking into account symmetries in 
nuclear systems, the Hamiltonian can be written as
\begin{equation}
H^{f}=\sum_a \epsilon_a \hat{n}_a+\sum_{a \leqslant b, c \leqslant d} \sum_{J T} V_{J T}(a b ; c d) \hat{T}_{J T}(a b ; c d),
\label{Eq:shell_model}
\end{equation}
where $\hat{n}_a$ denotes the occupation number for the orbital $a$ with quantum numbers 
$n_a,l_a,m_a$. 
The coefficients in this Hamiltonian are parametrized by fitting to the experimental data. The first term is 
a one-body operator and the second term is a two-body operator given by 
\begin{equation*}
\hat{T}_{J T}(a b ; c d)=\sum_{J_z T_z} A_{J J_z T T_z}^{\dagger}(a b) A_{J J_z T T_z}(c d),
\end{equation*}
where $(JJ_z)$ and $(TT_z)$ denote the coupled spin and isospin quantum numbers, respectively. And $A_{J J_z T T_z}^{\dagger}(a b)$ is given by
\begin{align*}
    &\hat{A}_{J J_z, T T_z}^{\dagger}(a b) \equiv\left[c_a^{\dagger} \times c_b^{\dagger}\right]_{J J_z, T T_z} = \\
&\sum_{m_a, m_b}\left(j_a m_a, j_b m_b \mid J J_z\right) \sum_{\mu_a, \mu_b}\left(\frac{1}{2} \mu_a, \frac{1}{2} \mu_b \mid T T_z\right)\\ &\hat{c}_{j_a m_a, \frac{1}{2} \mu_a}^{\dagger}
\hat{c}_{j_b m_b, \frac{1}{2} \mu_b}^{\dagger}.
\end{align*}
The antisymmetric and normalized two-body state is defined by 
\begin{equation*}
|a b ; J J_z\rangle \equiv \frac{1}{\sqrt{1+\delta_{a b}}} \hat{A}_{J J_z}^{\dagger}(a b)|v\rangle.
\end{equation*}
where $|v\rangle$ denotes the vacuum state. 

There are various SM Hamiltonians that are applicable in different regions of the nuclide chart. In this paper, we focus on two SM parametrizations to demonstrate our method. The first one is the Cohen-Kurath SM~\cite{cohen1965effective}, which considers ${}^4 \mathrm{He}$ as an inert core with the valence nucleons filling the $p$-shell. The $p$-shell has the capacity of 12 nucleons, including 6 neutrons and 6 protons. Therefore, this model applies to nuclei from $^{4}$He to ${}^{16}\mathrm{O}$.
The second SM we consider is the Wildental SM, also known as the USD SM~\cite{brown2006new,preedom1972shell,fiase1988effective}, which treats the magic number nucleus ${}^{16}$O as the frozen core.
In this model, the valence nucleons can be filled into the $s$-$d$ shell with the capacity of 24 nucleons.
Thus this SM can describe the nuclei from $^{16}$O to $^{40}$Ca.
The applications of our method to other SMs are similar.

\section{The classical shadow}
\label{sec:Cl}
Quantum state tomography (QST) is widely used to characterize unknown quantum states, but it has two major drawbacks. Firstly, classical computers are inefficient at processing quantum states obtained through this method. Secondly, the resources required for QST increase linearly with the dimension of the Hilbert space~\cite{haah2016sample}. In many cases, we don't need to fully characterize a quantum state. Instead, we only need to know certain properties of the state. Performing randomized measurements can be efficient in such situations. One method is by using classical shadows.

The concept of shadow tomography was first introduced in Ref.~\cite{aaronson2018shadow}. The authors showed that using logarithmically scaled measurements ($\log^4(N)$) is sufficient to derive $N$ linear functions of an unknown state. Ref. ~\cite{huang2020predicting} provides a straightforward approach to constructing a classical description of an unknown state, known as the classical shadow. This method involves randomly measuring stabilizer observations of quantum states and then processing the measurements on a classical computer.

The steps for constructing a classical shadow are as follows. 
\begin{itemize}
    \item Randomly insert global Clifford gates $C$ before measurement.
    \item Measure the final state in computational basis $Z$ and derive the binary vector $| \Vec{z} \rangle$. This and the step above runs on a quantum computer.
    \item Calculate the density matrix $C^{\dagger} |\Vec{z}\rangle\langle \Vec{z}|C$. This step and all subsequent steps are done on a classical computer.
    \item Repeat the steps above $N$ times and calculate the expectation $\rho = \mathbb{E}_{ C}C^{\dagger}|\Vec{z}\rangle\langle\Vec{z}|C$ on the classical computer.
    \item Transform the matrix with an operator \begin{equation}
        M(\rho) = (2^n + 1)\rho - I.
    \end{equation}

\end{itemize}

\section{Modified Subspace Diagonalization Method}
\label{sec:Mo}
QSD is a powerful tool for solving eigenvalue problems on a quantum computer. The main advantage of quantum methods over classical ones is that quantum computers can generate more complex states. There are three common ways to generate suitable quantum states: real-time evolution, imaginary-time evolution, and quantum power method. In all these methods, the general eigenvalue problem that needs to be solved is:

\begin{equation}
H^{s}\Vec{x} = ES\Vec{x}.
\label{eq:sub_0 equation}
\end{equation}
Here, $H^{s}_{ij} = \langle \psi_i|H|\psi_j\rangle$ and $S_{ij} = \langle\psi_i|\psi_j\rangle$. $\{|\psi_i\rangle\},i=1,\dots,m$ are the basis of the subspace with dimension $m$.  $H$ represents the Hamiltonian of the system that we want to solve, while $H^{s}$ is the effective Hamiltonian in the subspace.

Our algorithm focuses on the first method, real-time evolution, where the state is prepared as $|\psi_j\rangle = e^{-iHt_j}|\Bar{0}\rangle$, with $|\Bar{0}\rangle$ being the initial state having a finite overlap with the actual ground state. However, this approach often encounters a problem where some singular values of the matrix $S$ are very small. As a result, statistical fluctuations of the elements in the matrices $H^{s}$ and $S$ can have a significant impact on the results. This is mainly due to the uncorrelated shot noise of the measurements. To overcome this shortcoming, one natural solution is to use state tomography to measure these final states and then calculate $S$ and $H^{s}$. However, this method requires a considerable amount of resources when the number of qubits is large, making it impractical. With a density matrix, it is also difficult to calculate matrix multiplication due to the large dimension. In this work, we represent these final states $\{|\psi_i\rangle\}$ using classical shadows $\{\rho_i\}$. Consequently, the original quantum state-based subspace algorithm cannot be employed. To handle the density matrix more conveniently, for an arbitrary matrix $X = X_{i,j}|\psi_i\rangle\langle \psi_j|, X_{i,j}\in \mathbb{C}$, we vectorize it as:
\begin{equation}
    X = \sum_{i,j} X_{i,j}|\psi_i\rangle\langle \psi_j|\longrightarrow |X\rangle\rangle = 
 \sum_{i,j}X_{i,j}|\psi_i\rangle\otimes |\psi_j\rangle.
    \label{eq: vectorize}
\end{equation}
Note that this is not the only way to vectorize. Another method commonly used in quantum information theory is the Pauli transfer matrix representation. In our work, it is more convenient to use the method of Eq.~\ref{eq: vectorize}.

A general map applied to $X$ can then be expressed as a matrix,
\begin{equation}
    AXB \longrightarrow A\otimes B^{T} |X\rangle\rangle.
\end{equation}
The inner product of two arbitrary vectors $|X_1\rangle\rangle, |X_2\rangle\rangle$ is defined by
\begin{equation}
    \langle \langle X_1|X_2\rangle\rangle = Tr(X_1^{\dagger}X_2).
\end{equation}
Instead of solving the stationary Schrödinger equation, we choose to solve a different eigenvalue problem:
\begin{equation}
    H\otimes I |\rho\rangle\rangle = E|\rho\rangle\rangle.
\end{equation}
where we use $I$ to denote the identity matrix with the same dimension as the density matrix. $|\rho\rangle\rangle$ and $E$ are eigenvector and eigenvalue, respectively.
The eigenstates of $H\otimes I$ are all highly degenerate, and the lowest eigenvalue is exactly $E_0$. The subspace we use to find the lowest eigenvalues is
\begin{align}
    \mathbb{V} = \{ \Sigma_{i,j}c_{i,j}|\rho_i \rho_j\rangle\rangle | c_{i,j}\in \mathbb{C}\},
\end{align}
using these classical shadows $\{\rho_i\}$. The specific steps are as follows
\begin{enumerate}
    \item Prepare an initial state $|\Bar{0}\rangle$ which has an overlap with the exact ground state.
    \item Construct a set consisting of $m$ different times ${t_j}$ for $j = 1,2,..,m$.
    \item Evolve the initial state under the system's Hamiltonian and derive the final state $|\psi_j\rangle = e^{-iHt_j}|\Bar{0}\rangle$.
    \item For each final state $|\psi_j\rangle$ ($j = 1,2,...,m$), construct its classical shadow $\{\rho_j\}$ using the algorithm presented in last section.
    \item Construct the set $\{\rho_i\rho_j\}$ for $i,j = 1,2,...,m$ consisting of the basis of matrix space and relabel the term of the set by $\sigma_i$ for $i=1,2,...,m^2$.
    \item Calculate the matrices $\tilde{S}$ and $\tilde{H}^{s}$ given by 
    \begin{equation}
        \tilde{S}_{i,j} = \langle\langle \sigma_i|\sigma_j\rangle\rangle,
    \end{equation}
    \begin{equation}
        \tilde{H}^{s}_{i,j} = \langle \langle \sigma_i|H\otimes I|\sigma_j\rangle\rangle.
    \end{equation}
    \item Solve the eigenvalue problem 
    \begin{equation}
        \tilde{H}^{s}\vec{c} = E\tilde{S}\vec{c}.
        \label{eq:sub equation}
    \end{equation}
\end{enumerate}
Here $\tilde{H}^{s}$ is the effective Hamiltonian defined in the subspace $\mathbb{V}$.
We show in the following theorems that the lowest possible eigenvalue solved with Eq.~\ref{eq:sub equation} is the ground state energy of $H$.

\begin{theo}
    If an infinite number of shots are used in constructing the classical shadow, the eigenvalues of Eq.~\ref{eq:sub equation} and Eq.~\ref{eq:sub_0 equation} are the same.
\end{theo}
 We first consider using $\rho_i\rho_{j=1}$ for $i=1,2,3,\dots,m$.
As the shots tend to infinity, the classical shadow becomes an exact density matrix, which we denote as
$\rho_{i} = |\psi_i\rangle\langle\psi_i |$. At the same time, the product of the two classical shadows becomes
\begin{equation}
    \rho_i \rho_1 = |\psi_i\rangle\langle\psi_i|\psi_1\rangle\langle\psi_1 |.
\end{equation}
Its vectorized form is
\begin{equation}
    |\rho_i\rho_1\rangle\rangle  = \langle\psi_i|\psi_1\rangle |\psi_i\rangle\otimes |\psi_1\rangle.
\end{equation}
Suppose $\vec{x}$ is one solution of Eq.~\ref{eq:sub_0 equation} with the energy $E_a$, i.e., 
\begin{equation}
    \Sigma_j \langle\psi_i|H|\psi_j\rangle x_j =  \Sigma_j E_a \langle\psi_i|\psi_j\rangle x_j .
\end{equation}
We can construct the general eigenvector from the solution:
\begin{align}
    &\Sigma_j\langle \langle \rho_i\rho_1|H\otimes I|\rho_j\rho_1\rangle\rangle x_j/\langle\psi_j|\psi_1\rangle\nonumber \\ 
      = &\Sigma_j \langle\psi_1|\psi_i\rangle\langle\psi_j|\psi_1\rangle (\langle\psi_i| \langle\psi_1|)|H\otimes I(|\psi_j\rangle|\psi_1\rangle) x_j/\langle\psi_j|\psi_1\rangle\nonumber\\
     = &\Sigma_j \langle\psi_1|\psi_i\rangle \langle\psi_i|H|\psi_j\rangle x_j\nonumber\\
     = & \Sigma_j \langle\psi_1|\psi_i\rangle E_a \langle\psi_i|\psi_j\rangle x_j\nonumber \\
     =& \Sigma_j \langle\psi_1|\psi_i\rangle\langle\psi_j|\psi_1\rangle E_a \langle\psi_i|\psi_j\rangle x_j/\langle\psi_j|\psi_1\rangle\nonumber\\
     =& E_a\Sigma_j \langle\langle \rho_i\rho_1|\rho_j\rho_1\rangle\rangle x_j/\langle\psi_j|\psi_1\rangle.
\end{align}
Thus $\vec{x}^{\prime}$ where $x^{\prime}_{i} = x_i/\langle\psi_j|\psi_1\rangle$ is the solution of Eq.~\ref{eq:sub equation} and the corresponding eigenvalue is $E_a$. There are $m$ degenerate eigenvectors, each has the form 
$\Sigma_i x^{\prime}_i|\rho_i\rho_j\rangle\rangle$, for $j=1,2,\dots,m$. Since Eq.~\ref{eq:sub_0 equation} has $m$ different eigenvectors, all $m^2$ eigenvectors of Eq.~\ref{eq:sub equation} can be constructed and the spectrum of two equations are exactly the same.

\begin{theo}
    Suppose $E_0$ is the exact ground state energy.
    The eigenvalues $E_s$ obtained through the above process always satisfy $E_s \geq  E_0$. 
    When the number of shots tends to infinity and $\mathbb{V}$ exactly covers the density matrix of the ground state, the equal sign can be obtained.
\end{theo} 
The proof is straightforward. 
For any solution $\Vec{c}$ of Eq.~\ref{eq:sub equation}, the following formulas are established for the corresponding vector $|\rho(\Vec{c})\rangle\rangle$ in $\mathbb{V}$:
\begin{equation}
    \frac{\langle\langle \rho(\vec{c})|H\otimes I|\rho(\Vec{c})\rangle\rangle}{\langle\langle \rho(\Vec{c})|\rho(\vec{c})\rangle\rangle} \ge E_0.
\end{equation}
The key to this fact is that we use a quantum computer to generate a set of basis vectors $|\sigma_i\rangle\rangle$, instead of directly using a quantum computer to generate the $S$ and $H^{s}$ matrices. Since the basis vectors are stored in the classical computer, the construction of the $\tilde{S}$ and $\tilde{H}^{s}$ matrices is accurate.

Although a finite number of shots will not cause the calculated energy value to be less than the exact ground state energy, it will still cause deviation and fluctuations in the $\Tilde{H}^s$ and $\Tilde{S}$ matrix elements. We use the following two theorems to establish the relationship between the deviation and fluctuation of matrix elements and the number of shots.

\begin{theo}
    Assume that we measure each quantum state $M$ times to construct its shadow $\tilde{\rho}_i$, for $i=1,2,\dots,m$, the expected deviation $\mathbf{E}[\langle\langle\rho_j\rho_i|H\otimes I|\rho_l\rho_m\rangle\rangle - \langle\langle\Tilde{\rho}_j\Tilde{\rho}_i|H\otimes I|\Tilde{\rho}_l\Tilde{\rho}_m\rangle\rangle ]$ is $\sim \mathcal{O}({d^3}/{M^2})$.
    \label{theo: bias}
\end{theo}
The detailed proof is given in Appendix~\ref{proof: bias}. With the finite number of measurements, our estimate of the matrix elements of $\Tilde{H}^s$ and $S$ has a fluctuation range, which can be estimated by the variance. The following theorem presents the relationship between the variance and the dimension of the Hilbert space $d$ and the number of measurements $M$.
\begin{theo}
    Assume that we measure each quantum state $M$ times to construct its shadow $\tilde{\rho}_i$, the expected variance is
    \begin{align}
    &\mathbf{E}\left[\text{Var}[| 
 \langle\langle\rho_j\rho_i|H\otimes I|\rho_l\rho_m\rangle\rangle - \langle\langle\Tilde{\rho}_j\Tilde{\rho}_i|\otimes I|\Tilde{\rho}_l\Tilde{\rho}_m\rangle\rangle |]\right]\nonumber\\
 =&\sum_{a,b}c_{a,b}\frac{d^a}{M^b},
    \end{align}
    where $a/b\leq 1$.
    \label{theo: var}
\end{theo}
See the proof in Appendix ~\ref{proof: var}.
This result shows that although our algorithm requires multiplying four shadows, the number of measurements required to construct each shadow only increases linearly with the Hilbert space dimension. This is consistent with the number of measurements required to simply multiply two shadows~\cite{huang2020predicting}.

 Besides, we show later by numerical 
simulations that this method also retains the advantages of the original method, 
that is, the accuracy of the calculation increases exponentially as the subspace becomes larger.



\section{The full algorithm}
\label{sec:Fu}
    In this section, we describe the full algorithm. In the NISQ era, we want to reduce the number of qubits as much as possible. Therefore, our first step is to simplify the original Hamiltonian $H^{f}$ of nuclear SM. When calculating the ground state of a given nucleus, we don't need to use the entire Hilbert space but only need to solve the problem in the subspace with a corresponding number of particles. Considering only subspaces with a given number of particles can help reduce the number of qubits required as we can use the qubit-efficient encoding method to encode the fermionic states into qubit states~\cite{shee2022qubit}. Suppose $H$ is a matrix representation of the Hamiltonian $H^f$ in the subspace. Using the algebraic relation of fermionic operators, the matrix elements of $H$ can be easily calculated as follows: 
    \begin{enumerate}
        \item Assign a number to represent each occupiable orbital in the SM.
        \item Generate a set of occupied states $\{a_{k_1}^{\dagger}a_{k_2}^{\dagger}a_{k_3}^{\dagger}\dots a_{k_p}^{\dagger}|v\rangle\}$,
        where $p$ is the number of valence nucleons of the nucleus to be studied. $|v\rangle$ represents the vacuum state.
        To ensure that these states are linearly independent, we specify $k_1> k_2 > k_3 >\dots > k_p$. Without loss of generality, we specify that $a_{k_1}^{\dagger}a_{k_2}^{\dagger}a_{k_3}^{\dagger}\dots a_{k_p}^{\dagger}|v\rangle$ is the $k$th states in the set. 
        
        \item Calculate the matrix elements $H_{k,k^{\prime}} = \langle v|a_{k_p^{'}}\dots a_{k_1^{'}} H^{f} a_{k_1}^{\dagger}\dots a_{k_p}^{\dagger}|v\rangle$. 
        A classical computer can efficiently compute this quantity by exploiting the algebraic relationship of $a$ and $a^{\dagger}$.
    \end{enumerate}
    We can verify that such a Hamiltonian is sparse. We can see from Eq.~\ref{Eq:shell_model} that the Hamiltonian can be written as 
    \begin{equation}
        H^{f} = \Sigma_i g_i H_i,
    \end{equation}
    where $H_i$ can be expressed by $H_i = a_{i_1}^{\dagger}a_{i_2}^{\dagger}a_{i_3}a_{i_4} := a^{\dagger}_{\vec{i}_{12}}a_{\vec{i  }_{34}}$, where $\vec{i}_{12} = (i_1,i_2)$ and $\vec{i}_{34} = (i_3,i_4)$. With an $n$-dimension vector $\Vec{x}$, we use $a^{\dagger}_{\vec{x}}$ to denote the operator $a^{\dagger}_{x_1}a^{\dagger}_{x_2}\dots a^{\dagger}_{x_n}$.
    We only need to verify that the matrix $H_i$ has one non-zero term in each column. 
    Owing to the algebraic relationship of the fermionic operator:
    \begin{equation}
        H_i a_{\vec{k}}^{\dagger} |v\rangle = h(\vec{i}_{12},\vec{i}_{34},\vec{k})a_{\vec{k}^{\prime}}^{\dagger} |v\rangle,
    \end{equation}
    where $\Vec{k}^{\prime}$ is derived by replacing $i_3$ and $i_4$ in $\vec{k}$ with $i_1$ and $i_2$. The function $h$ equals $0$ or $\pm 1$. When $i_3$ and $i_4$ are contained in $\vec{k}$, $h = \pm 1$; the sign depends on the relationship of the elements in $\vec{i}_{12}$ and $\vec{i}_{34}$. Otherwise $h = 0$. 
    The only one annihilation operator series $a_{\vec{k}^{''}}$ that satisfies $\langle v | a_{\vec{k}^{''}} a_{\vec{k}^{\prime}}^{\dagger} | v\rangle \neq 0$ is the case that
    $\vec{k}^{''}$ and $\vec{k}^{\prime}$ have the same elements. Thus each column of the $H_i$ matrix has at most one non-zero element. As a result, several existing methods can be employed to achieve the evolution of this Hamiltonian~\cite{low2017optimal,berry2007efficient,berry2014exponential}. If $H$ is a $d\times d$ matrix, then only $ n = \lceil\log_2(d)\rceil$ qubits are needed.

It is worth noting that, in addition to this encoding method, other encoding methods are available, such as the Jordan-Wigner transformation and the gray code. The former is relatively straightforward and widely used since it is easy to realize the evolution of the Hamiltonian. The latter requires fewer qubits than the former and has gained attention in recent years~\cite{sawaya2020resource,di2021improving,siwach2021quantum}.

On a quantum computer, the first step is to initialize the qubits and prepare a state with a finite overlap with the exact ground state $\rho_0$. In our numerical experiments, we always set the initial state of $n$ qubits to be $|0\rangle^{\otimes n}$. When the nucleus becomes quite large, it may be necessary to design the initial state more carefully. Next, a set of discrete times $t_i,i=1,2,\dots,m$ is chosen. It's worth noting that the time difference between $t$ need not be very large to overcome the error caused by shot noise. Then, we evolve the initial state $\rho_{ini} = |0\rangle\langle 0|^{\otimes n}$ under the system's Hamiltonian, $\tilde{\rho}_i = U_i \rho_{ini} U_i^{\dagger}$, where $U_i = e^{-iHt_i}$. The Clifford randomized measurements are chosen to construct the classical shadow $\rho_i$ for different $i$. Given the classical shadows, we define the shadow space $\mathbb{V}$, which is spanned by $|\rho_i\rho_j\rangle\rangle$. Finally, we calculate the matrices $\tilde{H}^{s}$ and $\tilde{S}$ and solve the general eigenvalue problem. Recording the information of the classical shadows and the construction of $\tilde{H}^{s}$ and $\Tilde{S}$ can be effectively done by classical computers. 
    We summarize the steps as follows:
    \begin{enumerate}
        \item Construct the reduced Hamiltonian $H$ from the full Hamiltonian.
        \item Choose a set of discrete time $\{t_i\}$ for $i=1,...,n$.
        \item For each time $t_i$, evolve the initial state $|0\rangle^{\otimes m}$ by $e^{-iH t_i}$ and get the final state $|\psi_i\rangle$.
        \item For each final state, randomly choose a global Clifford gate $C_j$ and apply $C_j$ to the state.
        \item  Apply $Z$ measurement on the state $C_j|\psi_i\rangle\langle\psi_i|C_j^{\dagger}$, obtain a binary vector $|b\rangle$.
        \item Repeat step 4 and step 5 $N$ times, and construct the classical shadow ${\rho}_i$ for the final state $|\psi_i\rangle$.
        \item Construct the subspace comprising the vectors $|\rho_i \rho_j\rangle\rangle$, and then calculate the effective Hamiltonian $\tilde{H}^{s}$ and overlap matrices $\tilde{S}$ in this subspace.
        \item Solve the general eigenvalue problem $\tilde{H}^{s}\vec{c} = E\tilde{S}\vec{c}$.
    \end{enumerate}

\section{Numerical Simulations}
\label{sec:Si}

To test the effectiveness of this algorithm, we conducted numerical experiments to calculate the ground state of six nuclei: ${}^6\mathrm{He},{}^6\mathrm{Li},{}^7\mathrm{Bo},{}^8\mathrm{Li},{}^{18}\mathrm{O},$ and ${}^{18}\mathrm{F}$. The Cohen-Kurath SM was used to describe the first four nuclei, while the USD SM was used for the last two.

\begin{figure*}[ht!]
\label{F:Fig}
    \subfigure[Data of ${}^{8}$Li]{\includegraphics[width=0.3\linewidth]{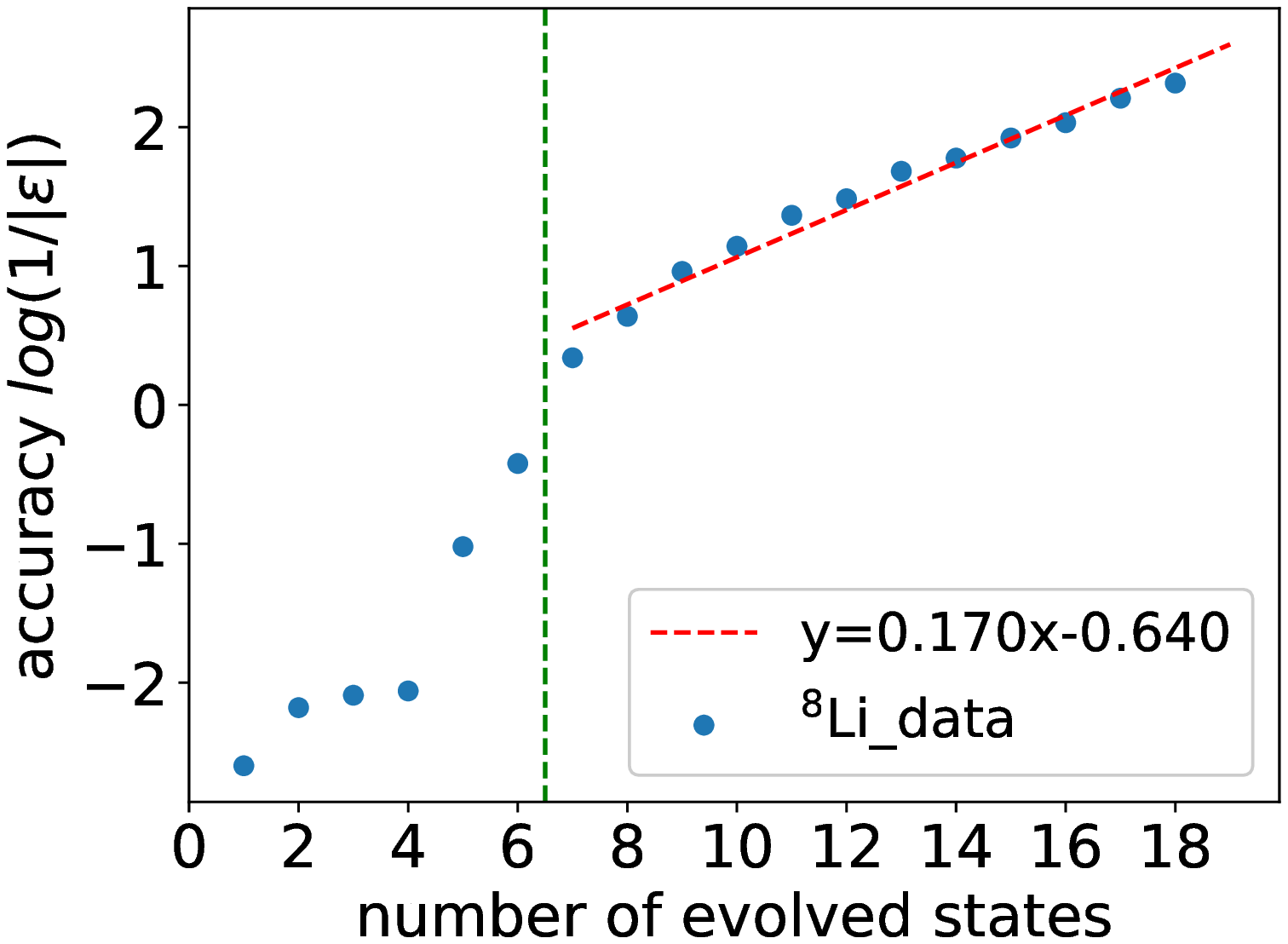}\label{fig:Li8_diff}}
    \subfigure[Data of ${}^{18}$F]{\includegraphics[width=0.3\linewidth]{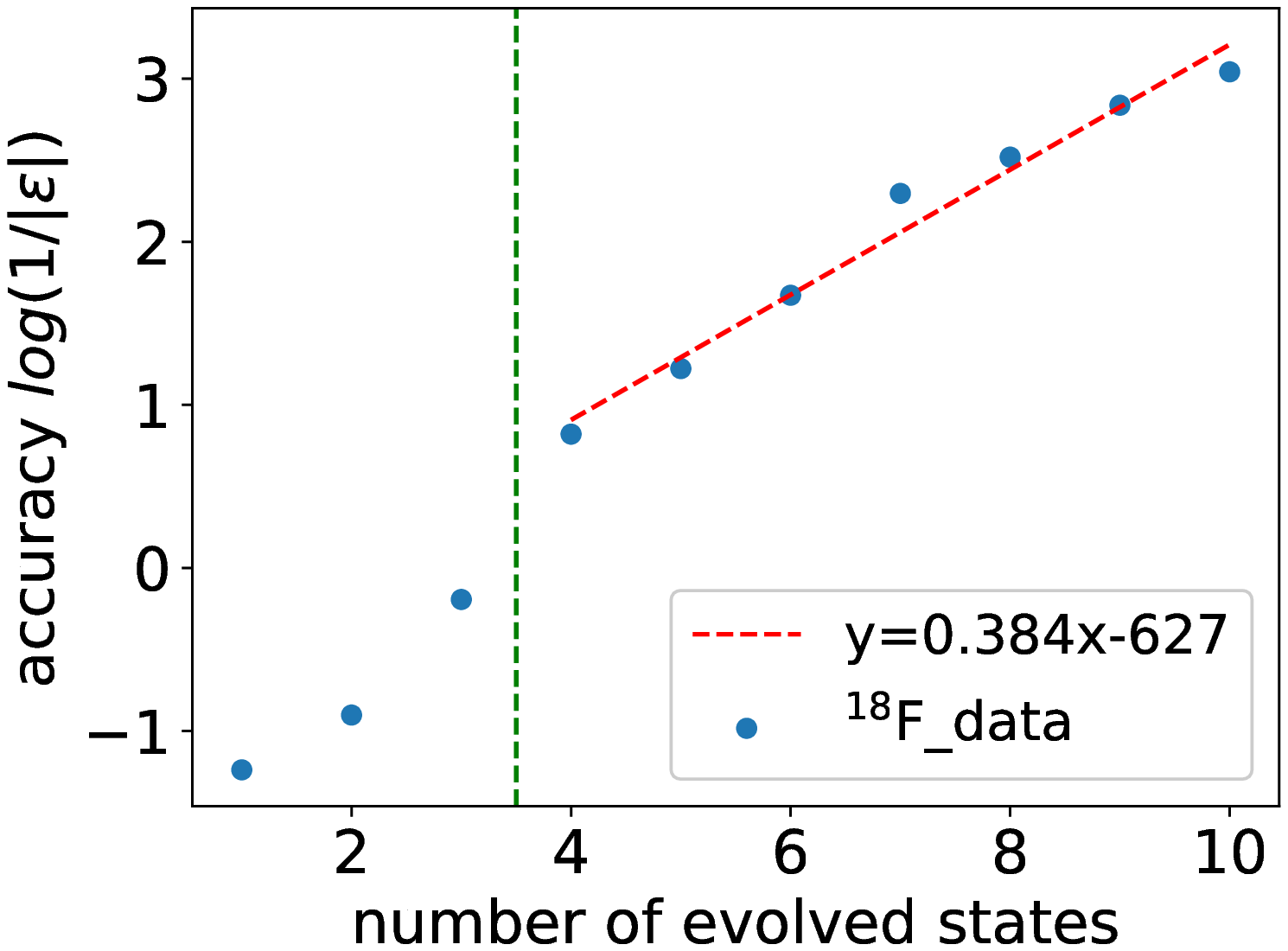}\label{fig:F18_diff}}\\
    \subfigure[Data of ${}^{6}$He]{\includegraphics[width=0.3\linewidth]{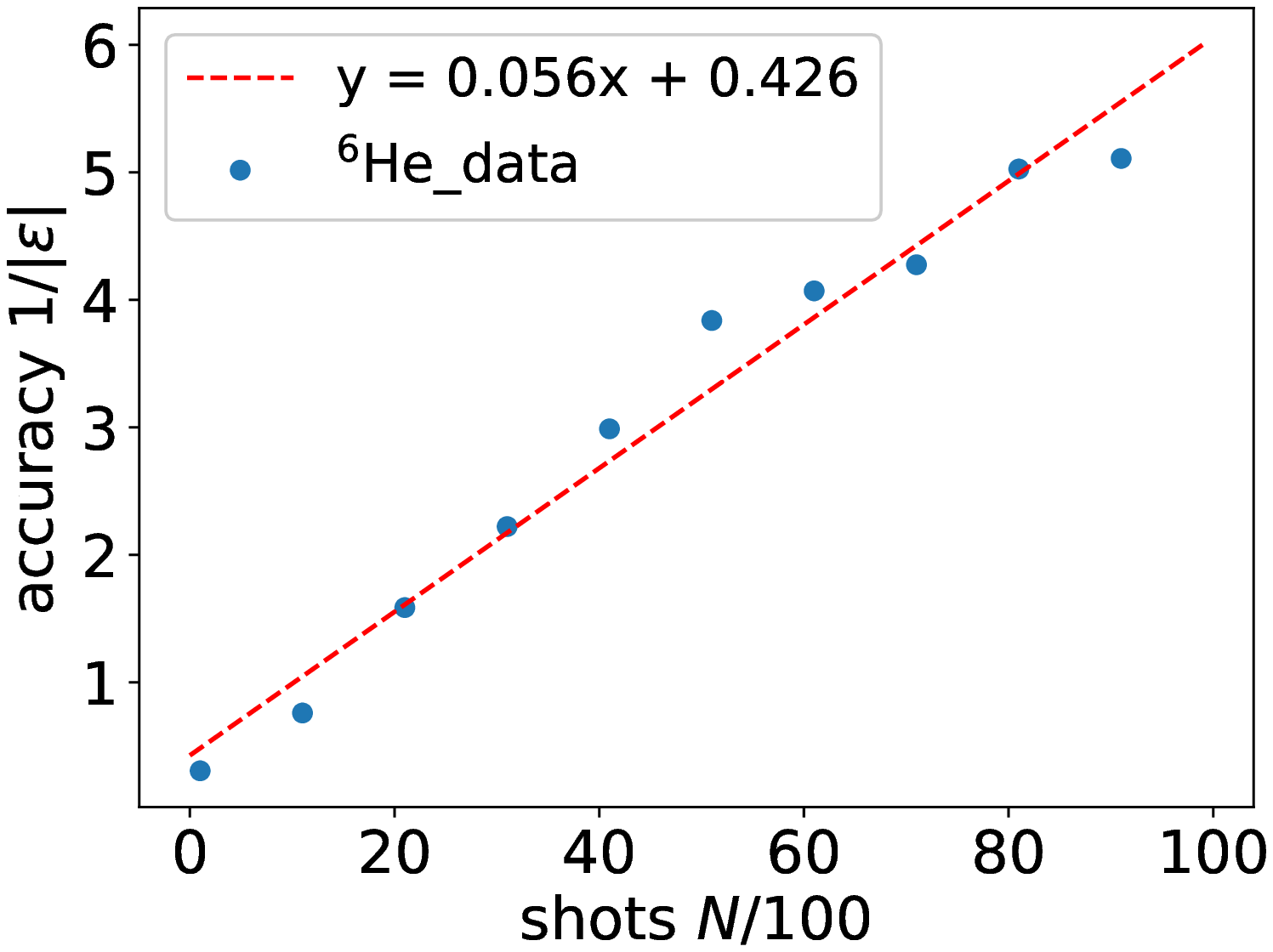}\label{fig:He6}}
    \subfigure[Data of ${}^{6}$Li]{\includegraphics[width=0.3\linewidth]{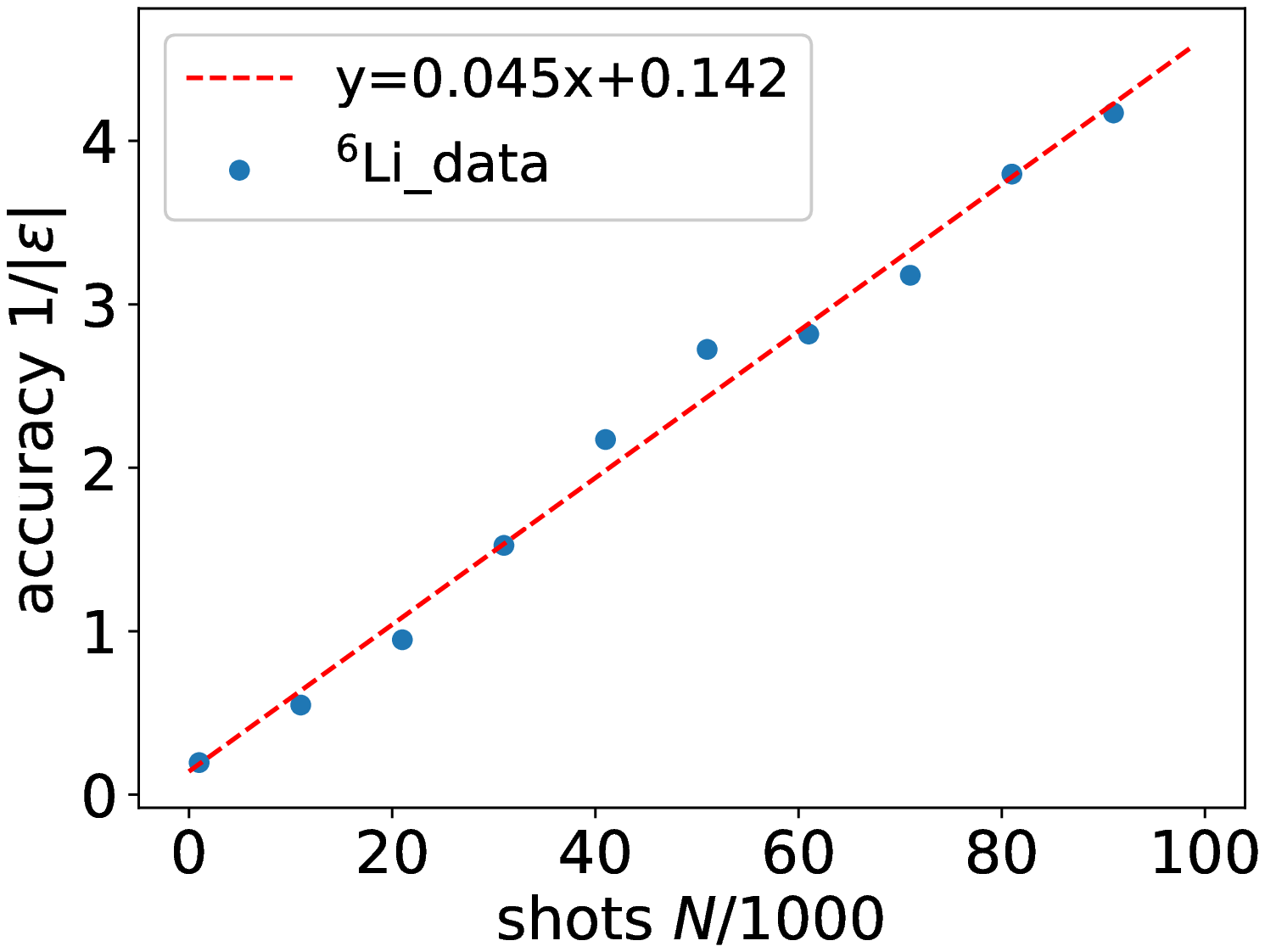}\label{fig:Li6}}
    \subfigure[Data of ${}^7$Bo]{\includegraphics[width=0.3\linewidth]{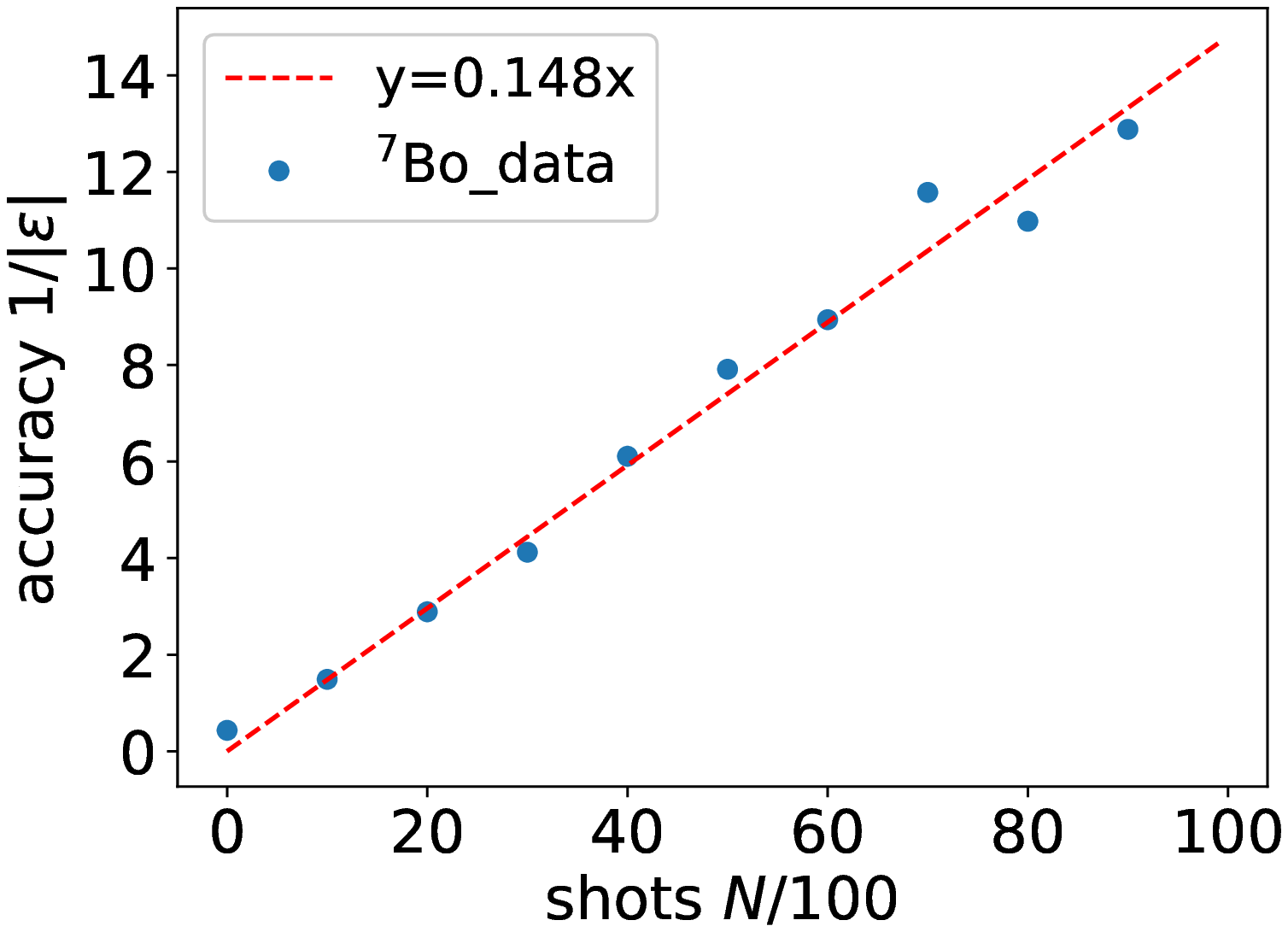}\label{fig:Bo7}}
   \subfigure[Data of ${}^{8}$Li]{\includegraphics[width=0.3\linewidth]{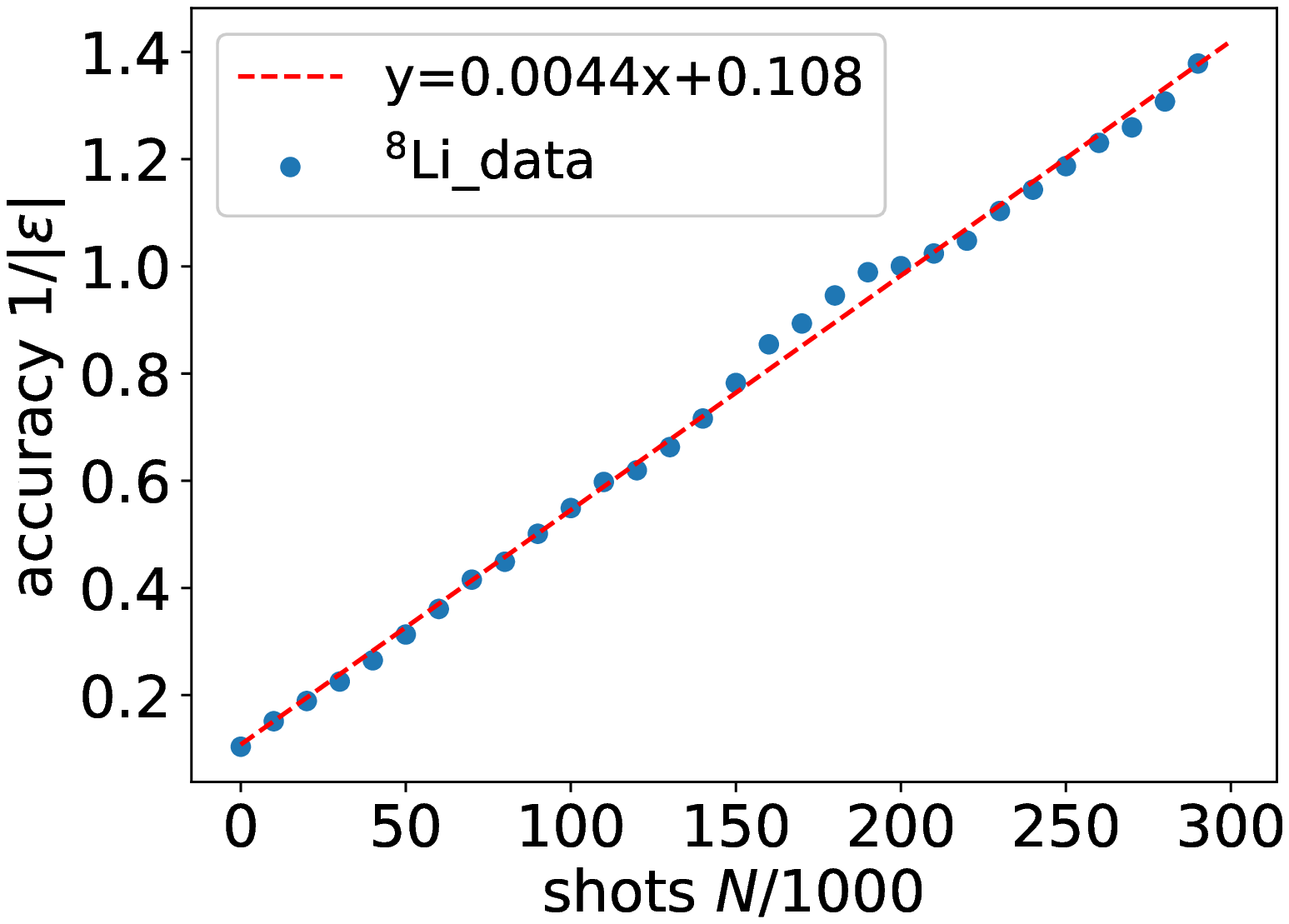}\label{fig:Li8}}
    \subfigure[Data of ${}^{18}$F]{\includegraphics[width=0.3\linewidth]{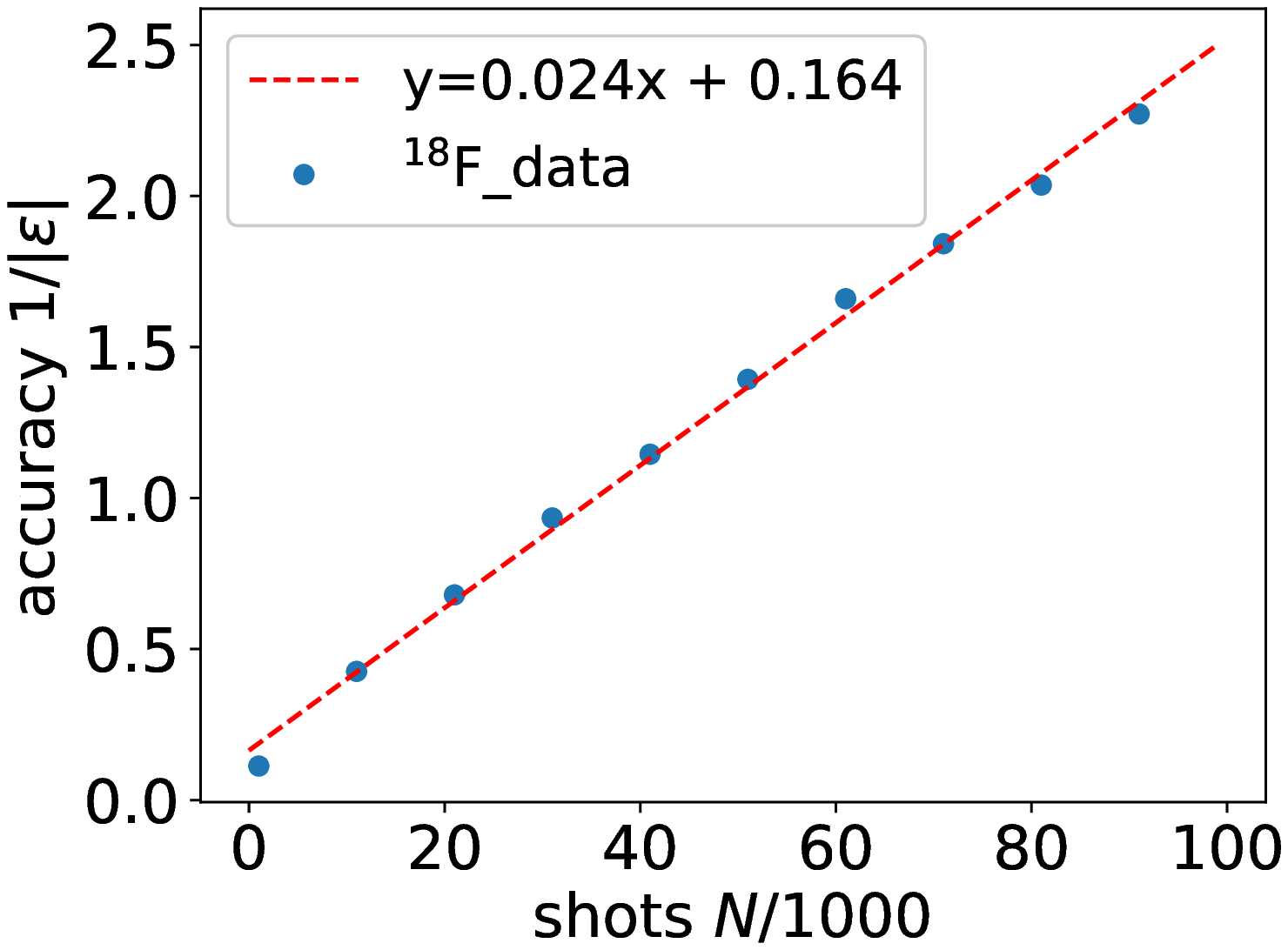}\label{fig:F18}}
    \subfigure[Data of ${}^{18}$O]{\includegraphics[width=0.3\linewidth]{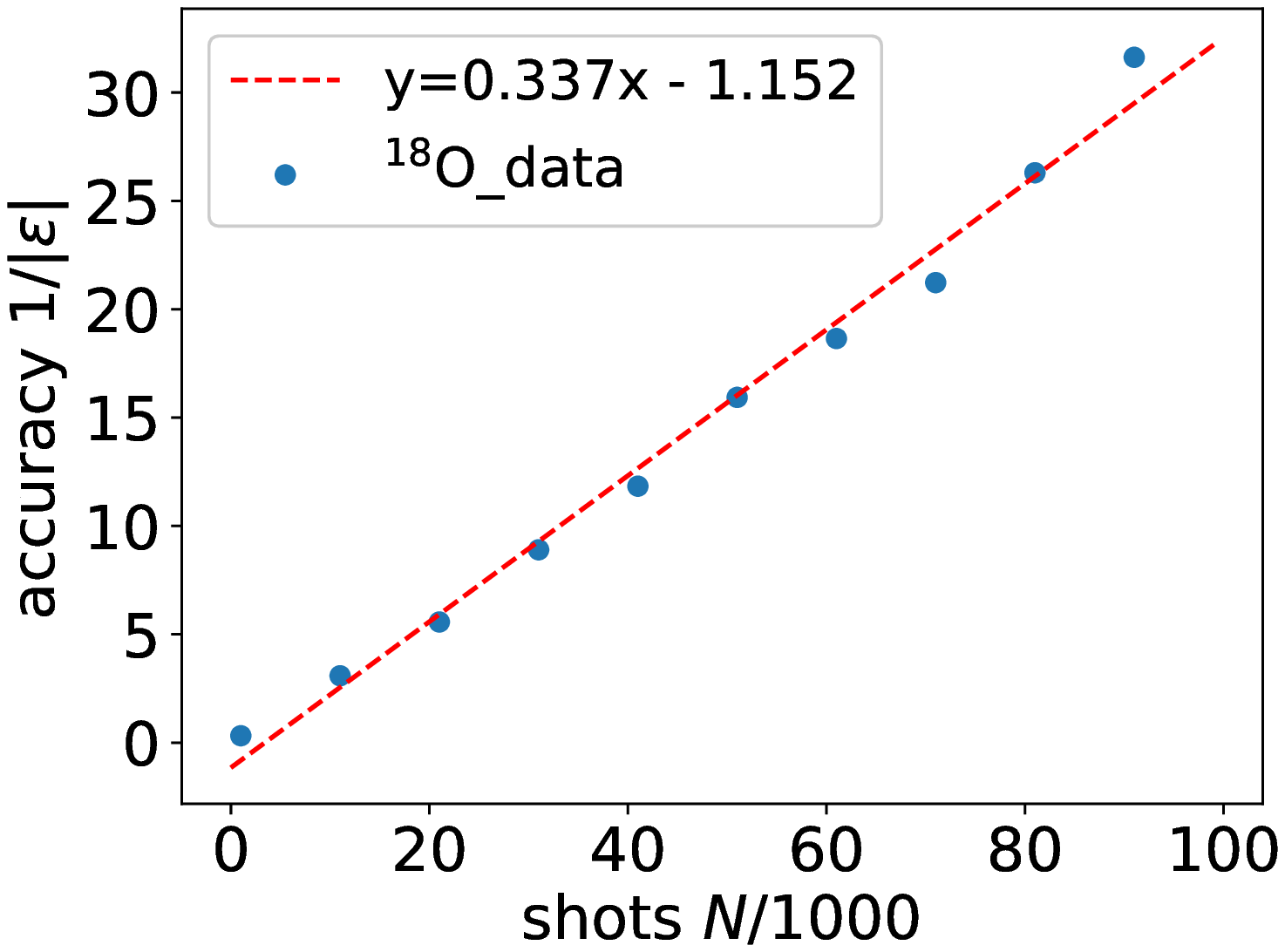}\label{fig:O18}}

\caption{The accuracy as a function of the number of evolved states and the number of shots. The blue dots represent the calculated results, while the red line is a straight line that is fitted to the data. 
The error here, in a unit of MeV, is defined as the absolute value of the difference between the calculated result and the ideal value derived by directly diagonalizing the Hamiltonian of the SM.
Figures \ref{fig:Li8_diff} and \ref{fig:F18_diff} illustrate how the accuracy varies with the number of evolved states. The horizontal axis represents the number of evolved states, while the vertical axis shows the logarithm of the reciprocal of the error. Fig.~\ref{fig:He6} $\sim$ Fig.~\ref{fig:O18} characterize the relation between the error and the number of shots for six nuclei. The horizontal axis represents the number of shots, while the vertical axis shows the reciprocal of the error.}
\end{figure*}

Table~\ref{table:data} presents some computational data for these six nuclei. First, ${\bf MNES} = M$ denotes the minimum number of evolved states needed such that the ground state is a linear combination of the evolved states ${e^{-iHx}|0\rangle^{\otimes n}}$, where $x = 1,2,\dots,M$. Generally, the accuracy of the calculated result increases with the number of evolved states in the subspace. Here we use $M$ number of states in order for better comparison between different nuclei. In addition, we determine the required number of qubits for each nucleus. For instance, ${}^6\mathrm{Li}$ has one neutron and one proton distributed in 6 neutron orbitals and 6 proton orbitals, respectively. Therefore, its wave function must be represented by $6\times 6 = 36$ basis states. Thus, $\lceil\log_2(36)\rceil = 6$ qubits are sufficient. Finally, the discrepancy between the calculated and experimental ground state energy is presented using the fixed shots and $M$ evolved states.

 \begin{table}[]
    \begin{tabular}{|l|l|l|l|l|l|l|}
    \hline
    Nucleus        & ${}^6\mathrm{He}$                   & ${}^6\mathrm{Li}$                   & ${}^7\mathrm{Bo}$ & ${}^8\mathrm{Li}$ & ${}^{18}\mathrm{F}$                   & ${}^{18}\mathrm{O}$                   \\ \hline
    MNES        & 2                     & 4                     & 3                        & 7                        & 4                     & 3                     \\ \hline
    Qubits          & 4                     & 6                     & 5                        & 7                        & 8                     & 6                     \\ \hline
    Shots           & $10^4$                  & $10^5$              & $10^4$                     & $3\times10^5$           & $10^5$                     & $10^5$ \\ \hline
    Difference(MeV) & 0.1957                & 0.2398                & 0.0776                   & 0.8616                   & 0.4402                & 0.0316                \\ \hline
    \end{tabular}
    \caption{ Parameters and calculation results for different nuclei.
    {\bf MNES} means the minimum number of evolved states whose linear combination equals to the ground state.
    {\bf Qubits} denotes the number of qubits used in the simulation of quantum computing. {\bf Shots} refers to the number 
    of measurements taken to construct the classical shadow of each state. {\bf Difference} is defined as the absolute value of the difference between the calculated ground-state energy and the laboratory-measured energy from Ref.~\cite{fiase1988effective}.}
\label{table:data}
\end{table}

To explore the accuracy of ground state energy calculations for each nucleus as a function of the number of evolved states, we fix the number of shots used to construct each classical shadow and then increase the number of evolved states, starting from the minimum number. We choose to use ${}^8\mathrm{Li}$ and ${}^{18}\mathrm{F}$ as examples because the dimensions of their corresponding Hilbert spaces are much larger than the minimum number of evolved states required. This provides more flexibility to increase the subspace. The results are shown in Fig.~\ref{fig:Li8_diff} and Fig.~\ref{fig:F18_diff}. We define the accuracy as the inverse of error $\epsilon$ (in MeV), which is the absolute value of the difference between the calculated result and the ideal value derived by directly diagonalizing the Hamiltonian of SM. To aid in observation, here we use logarithmic coordinates for the vertical axis. The green dashed line marks the MNES for the two nuclei. As can be seen from the figure, the calculated results appear to roughly form a straight line after the number of evolved states is beyond the MNES; the data are fit with a red line using the least squares method. This result indicates that, once the minimum number is reached and the number of evolved states continues to increase, the error caused by the finite number of shots decreases exponentially. However, simply increasing the number of evolved states is not enough to achieve infinite accuracy with a fixed number of shots. To further increase the accuracy, we need to increase the number of shots. 


We show the effect of the number of shots in Figure~\ref{fig:He6} to~\ref{fig:O18}, where the minimum number of evolved states is used for each nucleus. In contrast to Fig.~\ref{fig:Li8_diff} and Fig.~\ref{fig:F18_diff}, the vertical axis in this case represents the inverse of the error, while the horizontal axis represents the number of shots. The red line is the fit of the data using the least squares method.


By analyzing the measurement data represented by the blue dots, we can observe that:
\begin{align*}
    \frac{1}{|\epsilon|} & \propto N,\\
    |\epsilon| &\propto \frac{1}{N}.
\end{align*}
This indicates that the accuracy $\frac{1}{|\epsilon|}$ of the result using our method is dependent on the number of shots $N$ and approaches the Heisenberg limit, which was not observed in the original subspace diagonalization method.

\begin{figure}[ht!]
\centering
\includegraphics[width=0.9\linewidth]{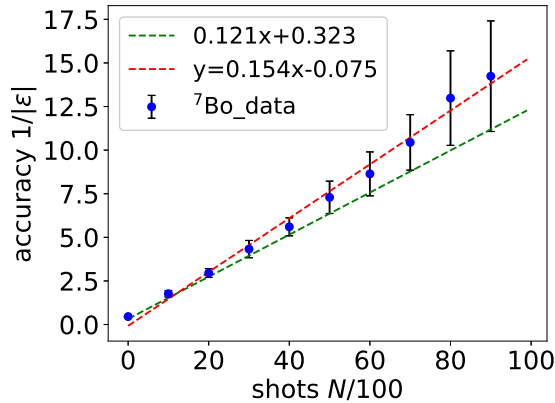}
\caption{The accuracy of the ground energy of ${}^7 \mathrm{Bo}$ as a function of the number of shots. The error bar is defined as the standard deviation of $1/|\epsilon|$. The green line marks the lower bound of the error bar.}
\label{fig:Bo7_bar}
\end{figure}

To further confirm this phenomenon, we performed calculations for ${}^7\mathrm{Bo}$ and calculated its error bar, defined as the standard deviation of $1/|\epsilon|$. We also examined how the lower bound of the error bar changed with the number of shots. As shown in Figure~\ref{fig:Bo7_bar}, the green line was fitted to the lower bound of the error bar. We see that the dependence remains close to a straight line, with the only difference being a slightly smaller slope. Therefore, we conclude that the Heisenberg limit dependence still holds.


\section{Conclusion and Outlook}
\label{sec:Co}
In this paper, we present a novel quantum algorithm for computing the ground state energy of nuclear systems. Our approach combines classical shadow techniques with the modified QSD method. The modified QSD method eliminates the need for Hadamard tests and requires fewer two-qubit gates than the original method, making it more suitable for NISQ devices. One of the distinctive characteristics of this method is that the relationship between accuracy and the number of shots follows the Heisenberg limit. This property can greatly reduce the number of shots needed in experiments. Additionally, this algorithm resolves the ill-conditioned overlap matrix problem that is often encountered in original algorithms.
It's worth noting that this method is not limited to nuclear systems and can be applied to a range of problems involving ground-state calculations.

\begin{acknowledgments}
We thank fruitful discussions with Jinniu Hu. This work is supported by National Natural Science Foundation of China (Grant No. 12225507, 12088101) and NSAF (Grant No. U2330201, U2330401 and U1930403).
\end{acknowledgments}


\bibliography{ref.bib}
\onecolumngrid
\appendix
\section{Proof of Theorem \ref{theo: bias} and \ref{theo: var}}
\subsection{Notation}
We use $\{\rho_i\}$ to be the set of the exact density matrices. Use $\{\rho_i,k_i
\}$ to represent the corresponding classical shadow constructed by a single measurement. Then the exact density matrix is the expectation of the classical shadow:
\begin{align}
    \rho_{i} = \mathbf{E}[\rho_{i,k_i}].
\end{align}
We use Clifford-type shadow, which is
\begin{align}
    \rho_{i,k_i} &= (d+1)X_{i,k_i} - I_d,\\
    X_{i,k_i} &= U_{i,k_i}^{\dagger} |i,k_i\rangle\langle i,k_i| U_{i,k_i},
\end{align}
where $U_{i,k_i}$ is the random Clifford gate and $|i,k_i\rangle\rangle$ is the corresponding measurement result. $I_d$ is the $d\times d$ identity matrix.
Assume that we measure each quantum state $M$ times to construct its shadow $\Tilde{\rho}_i$, then
\begin{align}
    \Tilde{\rho}_{i} =\frac{1}{M} \sum_{k_i=1}^M \rho_{i,k_i}.
\end{align}
The error in energy estimation caused by a limited number of shots can be expressed as
\begin{align}
    &\langle\langle \rho_{i}\rho_{j}|H \otimes I|\rho_{l}\rho_{m}\rangle\rangle - \langle\langle \Tilde{\rho}_{i}\Tilde{\rho}_{j}|H \otimes I|\Tilde{\rho}_{l}\Tilde{\rho}_{m}\rangle\rangle\nonumber\\& = \langle\langle\Delta (\rho_i\rho_j)|H\otimes I|\Delta (\rho_l\rho_m)\rangle\rangle  \nonumber\\
    &+\langle\langle\Delta (\rho_i\rho_j)|H\otimes I|\rho_l\rho_m\rangle\rangle + \langle\langle\rho_i\rho_j|H\otimes I|\Delta (\rho_l\rho_m)\rangle\rangle,
    \label{eq: div}
\end{align}
where we have used the notation
\begin{align}
    \Delta(\rho_i\rho_j) = \rho_i\rho_j - \Tilde{\rho}_i\Tilde{\rho}_j.
\end{align}

\subsection{Proof of Theorem~\ref{theo: bias}}
\label{proof: bias}
In this subsection, we calculate the expected value of the error in the energy estimate. It is obvious that
\begin{align}
\textbf{E}[\Delta(\rho_i\rho_j)] = 0.
\end{align}
Then only the first term in Eq.~\ref{eq: div} contributes to the expectation. This term can be written as

\begin{align}
    &\mathbf{E}[\langle\langle\Delta (\rho_i\rho_j)|H\otimes I|\Delta (\rho_l\rho_m)]\rangle\rangle\nonumber\\ 
    & \leq \lambda_{max} \mathbf{E}[\langle\langle\Delta (\rho_i\rho_j)|\Delta (\rho_l\rho_m)\rangle\rangle]\nonumber\\
    & =\lambda_{max} ( \mathbf{E} [\langle\langle \Tilde{\rho}_{i}\Tilde{\rho}_{j}|\Tilde{\rho}_{l}\Tilde{\rho}_{m}\rangle\rangle] - \langle\langle \rho_{i}\rho_{j}|\rho_{l}\rho_{m}\rangle\rangle). 
    \label{eq: div bound}
\end{align}
When $i,j,l,m$ are not equal to each other, Eq.~\ref{eq: div bound} vanishes to zero. If $i=l$, we can expand Eq.~\ref{eq: div bound} as
\begin{align}
     \mathbf{E} [\langle\langle \Tilde{\rho}_{i}\Tilde{\rho}_{j}|\Tilde{\rho}_{i}\Tilde{\rho}_{m}\rangle\rangle] =\frac{1}{M^4} \sum_{k_i,k_j,k_l,k_m}^M\mathbf{E} [\langle\langle \rho_{i,k_i}\rho_{j,k_j}|\rho_{i,k_l}\rho_{m,k_m}\rangle\rangle].
     \label{eq: i=l}
\end{align}
Next, we need to discuss each case separately. Please note that we only need to consider the cross terms, i.e., at least two of $k_i$, $k_j$, $k_l$, and $k_m$ are equal. This is because when $k_i$, $k_j$, $k_l$, and $k_m$ are mutually unequal, they will be canceled by the second term of Eq~\ref{eq: div bound}, and thus will not contribute to Eq~\ref{eq: div bound}.
If $k_i = k_l$, then
\begin{align}
    &\mathbf{E} [\langle\langle \rho_{i,k_i}\rho_{j,k_j}|\rho_{i,k_i}\rho_{m,k_m}\rangle\rangle]\nonumber\\
    & = \mathbf{E} [\Tr(\rho_{j,k_j}\rho_{i,k_i}^2\rho_{m,k_m})]\nonumber\\
    & = \mathbf{E} [\Tr(\rho_{j}\rho_{i,k_i}^2\rho_{l})]\nonumber\\
    & = \Tr(\rho_j((d-1)\rho_i + dI)\rho_m)\nonumber\\
    & = (d-1)\Tr(\rho_j\rho_i\rho_m) + d(\rho_j\rho_m).
\end{align}
After traversing $M^4$ combinations, there are $M^3$ terms that meet this condition $k_i = k_l$. Multiplying this by the factor of $\frac{1}{M^4}$ in Eq.~\ref{eq: i=l}, the contribution of these terms to the expectation is $\mathcal{O}(d/M)$. 

Another case that needs to be discussed is when $i = m$ and $k_i = k_m$:
\begin{align}
    &\mathbf{E} [\langle\langle \rho_{i,k_i}\rho_{j,k_j}|\rho_{l,k_l}\rho_{i,k_i}\rangle\rangle]\nonumber\\
     = &\mathbf{E} [\Tr(\rho_{j,k_j}\rho_{i,k_i}\rho_{l,k_l}\rho_{i,k_i})]\nonumber\\
     = &\mathbf{E}[((d+1)^2\Tr(\rho_{j,k_j}X_{i,k_i}\rho_{l,k_l}X_{i,k_i}) - (d+1)\Tr(\rho_{j,k_j}X_{i,k_i}\rho_{l,k_l})\nonumber\\
    -&(d+1)\Tr(\rho_{j,k_j}\rho_{l,k_l}X_{i,k_i}) + \Tr(\rho_{j,k_j}\rho_{l,k_l}))]\nonumber\\
     = &\mathbf{E}[(d+1)^2\Tr(\rho_{j,k_j}X_{i,k_i}\rho_{l,k_l}X_{i,k_i})] - \Tr(\rho_{j,k_j}\rho_{i,k_i}\rho_{l,k_l})\nonumber \\ -& \Tr(\rho_{j,k_j}\rho_{l,k_l}\rho_{i,k_i}) - \Tr(\rho_{j,k_j}\rho_{l,k_l}).\label{eq: i=m}
\end{align}
Next, we focus on the first term of Eq.~\ref{eq: i=m}.
\begin{align}
    &\mathbf{E}[(d+1)^2\Tr(\rho_{j,k_j}X_{i,k_i}\rho_{l,k_l}X_{i,k_i})]\nonumber\\
     = &\mathbf{E}(d+1)^2\Tr(\rho_{j}X_{i,k_i}\rho_{l}X_{i,k_i})\nonumber\\
     =&\mathbf{E}[(d+1)^2 \Tr(\rho_j X_{i,k_i}) \Tr(\rho_l X_{i,k_i})]\nonumber\\
    \leq& (d+1)^2\sqrt{\mathbf{E}[|\Tr(\rho_j X_{i,k_i})|^2]\mathbf{E}[|\Tr(\rho_l X_{i,k_i})|^2]}.
\end{align}
Remark that the variance of estimating linear observations using classical shadow is~\cite{huang2020predicting} 
\begin{align}
    \mathbf{E}[|\Tr(\rho_j \rho_{i,k_i})|^2] - |\Tr(\rho_j \rho_{i})|^2 \sim \mathcal{O}(1).
\end{align}
Then we obtain
\begin{align}
    (d+1)^2\mathbf{E}[|\Tr(\rho_j X_{i,k_i})|^2] &=  \mathbf{E}[|\Tr(\rho_j \rho_{i,k_i} + \rho_j I_d)|^2]\nonumber\\
    &\sim \mathcal{O}(1).
\end{align}
Furthermore, it can be claimed that
\begin{align}
    &\mathbf{E}[(d+1)^2\Tr(\rho_{j,k_j}X_{i,k_i}\rho_{l,k_l}X_{i,k_i})]\nonumber\\
     = &(d+1)^2\mathbf{E}[\Tr(\rho_{j,k_j}X_{i,k_i}) \Tr(\rho_{l,k_l}X_{i,k_i})]\nonumber\\
     \leq & (d+1)^2 \sqrt{\mathbf{E}[|\Tr(\rho_{j,k_j}X_{i,k_i})|^2]}\sqrt{\mathbf{E}[|\Tr(\rho_{l,k_l}X_{i,k_i})|^2]}\nonumber\\
    \sim &\mathcal{O}(1).
\end{align}

Next, we discuss the case where $i = l$ and $j = m$:
\begin{align}
     \mathbf{E} [\langle\langle \Tilde{\rho}_{i}\Tilde{\rho}_{j}|\Tilde{\rho}_{i}\Tilde{\rho}_{j}\rangle\rangle] =\frac{1}{M^4} \sum_{k_i,k_j,k_l,k_m}^M\mathbf{E} [\langle\langle \rho_{i,k_i}\rho_{j,k_j}|\rho_{i,k_l}\rho_{j,k_m}\rangle\rangle].
\end{align}
In the case where $k_1 = k_3, k_2 = k_4$:
\begin{align}
    &    \mathbf{E} [\langle\langle \rho_{i,k_i}\rho_{j,k_j}|\rho_{i,k_i}\rho_{j,k_j}\rangle\rangle]
     = \mathbf{E} [\Tr(\rho_{j,k_j}^2\rho_{i,k_i}^2)]\nonumber\\
     = &\Tr(((d-1)\rho_j + dI_d)((d-1)\rho_i + dI_d))\nonumber\\
     \sim &O(d^3).
\end{align}
There are $M^2$ terms that meet this condition. Considering the factor of $1/M^4$, the contribution of these terms to the expectation is $\sim \mathcal{O}(d^3/M^2)$.

We also need to consider the case where $i=j=l\neq m$ and $k_i = k_j = k_l \neq k_m$.
\begin{align}
    &    \mathbf{E} [\langle\langle \rho_{i,k_i}\rho_{i,k_i}|\rho_{i,k_i}\rho_{m,k_m}\rangle\rangle]\nonumber\\
    = &\mathbf{E}[\Tr(\rho_{i,k_i}^3\rho_{m,k_m})]\nonumber\\
     =&\mathbf{E} [\Tr\left(((d^2-d-5)\rho_{i,k_i}+(d^2-d-4)I_d)\rho_{m,k_m} \right)]\nonumber\\
\sim &\mathcal{O}(d^2).  
\end{align}
There are $M^2$ items that meet this condition. Thus the contribution to the expectation is $\sim \mathcal{O}(d^2/M^2)$.

The final case we need to discuss is where $i = j = l = m$ and $k_i = k_j = k_l = k_m$:
\begin{align}
    &    \mathbf{E} [\langle\langle \rho_{i,k_i}\rho_{i,k_i}|\rho_{i,k_i}\rho_{i,k_i}\rangle\rangle]\nonumber\\
    = &\mathbf{E}[\Tr(\rho_{i,k_i}^4)]\nonumber\\
    =&\mathbf{E}[\Tr( (d^2-1)\rho_{i,k_i} + d^2 I_d)]\nonumber \\
    \sim & \mathcal{O}(d^3).
\end{align}
There are $M$ items that meet this condition. Thus the contribution to the expectation is $\sim \mathcal{O}(d^3/M^3)$.
Based on our discussion of the above four cases, we can assert that the expected deviation $E_s - E_0$ is $\sim \mathcal{O}({d^3}/{M^2})$
\subsection{Proof of Theorem~\ref{theo: var}}
\label{proof: var}
When considering variance, the situation becomes more complicated. We start with the relatively easy part first. Now, let's consider the variance of the second term of Eq.~\ref{eq: div bound}:
\begin{align}
    &\text{Var}[\langle\langle\Delta (\rho_i\rho_j)|H\otimes I|\rho_l\rho_m\rangle\rangle ]\nonumber\\  \leq& \lambda_{max}^2 \text{Var}[\langle\langle\Delta (\rho_i\rho_j)|\rho_l\rho_m\rangle\rangle ]\nonumber\\
    \leq& \lambda_{max}^2 \mathbf{E} [|\langle\langle\Delta (\rho_i\rho_j)|\rho_l\rho_m\rangle\rangle|^2]\nonumber\\
     = &\lambda_{max}^2 (\mathbf{E} [|\langle\langle \Tilde{\rho}_i\Tilde{\rho}_j|\rho_l\rho_m\rangle\rangle|^2] - |\langle\langle \rho_i\rho_j|\rho_l\rho_m\rangle\rangle|^2 )\nonumber\\
     =& \lambda_{max}^2(\mathbf{E}[|\Tr(\Tilde{\rho}_j\Tilde{\rho}_i\rho_l\rho_m)|^2] - |\Tr(\rho_j\rho_i\rho_l\rho_m)|^2).
\end{align}
Therefore, we can see that this is the variance of the quadratic nonlinear observation using shadow estimation. Thus we can assert that
\begin{align}
    \text{Var}[\langle\langle\Delta (\rho_i\rho_j)|H\otimes I|\rho_l\rho_m\rangle\rangle ]\sim\left({\frac{d^2}{M^2}}\right).
\end{align}
The variance of the third term of Eq.~\ref{eq: div bound} is the same as the second term. To calculate the variance of the first term, we need to use the following lemma.

\begin{lemma}
    For expressions of the form $\mathbf{E}[\Tr(X_{j,k_j}X_{i,k_i}X_{l,k_l}X_{m,k_m})\Tr(X_{j^{\prime},k_j^{\prime}}X_{i^{\prime},k_i^{\prime}}X_{l^{\prime},k_l^{\prime}}X_{m^{\prime},k_m^{\prime}})]$,  where the terms can be equal, we use $f$ to represent its degree of freedom, which means that there are $M^f$ terms that satisfy its conditions. If $f>0$ and at least one in $\mathcal{L} = \{(i,k_i), (j,k_j), (l,k_l), (m,k_m)\}$ is equal to one in $\mathcal{R} = \{(i^{\prime},k_i^{\prime}), (j^{\prime},k_j^{\prime}), (l^{\prime},k_l^{\prime}), (m^{\prime},k_m^{\prime})\}$, Then
    \begin{align}  &\mathbf{E}[\Tr(X_{j,k_j}X_{i,k_i}X_{l,k_l}X_{m,k_m})\Tr(X_{j^{\prime},k_j^{\prime}}X_{i^{\prime},k_i^{\prime}}X_{l^{\prime},k_l^{\prime}}X_{m^{\prime},k_m^{\prime}})]\nonumber\\& \sim \mathcal{O}\left(\frac{1}{d^f}\right).
    \end{align}
    \label{lemma1}
\end{lemma}

We use classification to prove the induction. Based on the equality relationship between the elements in $\mathcal{L}$ and $\mathcal{R}$, we can roughly divide them into five main categories, each with several subcategories. For each subcategory, we establish a series of equality relations that form the "skeleton". Elements outside the skeleton can be unequal, equal to adjacent elements, or equal to adjacent skeleton elements. In the following calculations, we assume that these elements are not equal to each other or to the skeleton elements. If one of these elements is equal to its neighbor, we can omit one of them from the calculation. This reduces the degree of freedom $f$ and the power of $d$ in the complexity by 1 each, without affecting the final conclusion.
We start with the simplest case. We use \textit{case 1.1} as an example to demonstrate this process.

\textbf{case 1}

One in $\{(i,k_i), (j,k_j), (l,k_l), (m,k_m)\}$ is equal to one in $\{(i^{\prime},k_i^{\prime}), (j^{\prime},k_j^{\prime}), (l^{\prime},k_l^{\prime}), (m^{\prime},k_m^{\prime})\}$. Without loss of generality, we assume $j = j^{\prime}$ and $k_j =k_j^{\prime} $. Now it can be divided into five cases.
\\
\textit{case 1.1} $k_j =k_j^{\prime} $.
    \begin{align}  &\mathbf{E}[\Tr(X_{j,k_j}X_{i,k_i}X_{l,k_l}X_{m,k_m})\Tr(X_{j,k_j}X_{i^{\prime},k_i^{\prime}}X_{l^{\prime},k_l^{\prime}}X_{m^{\prime},k_m^{\prime}})]\nonumber\\ = & 
    \frac{1}{(d+1)^6}\mathbf{E}[\Tr(X_{j,k_j}\rho_{i,k_i}\rho_{l,k_l}\rho_{m,k_m})\Tr(X_{j,k_j}\rho_{i^{\prime},k_i^{\prime}}\rho_{l^{\prime},k_l^{\prime}}\rho_{m^{\prime},k_m^{\prime}})]
    \nonumber\\
=&\frac{1}{(d+1)^6}\mathbf{E}[\Tr(X_\1\rho_{i}\rho_{l}\rho_{m})\Tr(X_\1\rho_{i}\rho_{l^{\prime}}\rho_{m^{\prime}})]\nonumber\\
\sim & \mathcal{O}\left(\frac{1}{(d+1)^8}\right).
    \end{align}

In \textit{case 1.1,} the skeleton is $k_j = k_j^{\prime}$, and we assume during the calculation that the other elements are not equal to each other, nor are they equal to $k_j$. If $k_j=k_i$, then we only need to delete $X_\2$ during the calculation process, and the final result is $\mathcal{O}(1/(d+1)^7)$. At the same time, the degree of freedom $f$ is reduced by 1.

\textit{case 1.2} $k_j =k_j^{\prime} = k_l $ and $k_m^{\prime} = k_i^{\prime}$. 
\begin{align}
&\mathbf{E}[\Tr(X_{j,k_j}X_{i,k_i}X_{j,k_j}X_{m,k_m})\Tr(X_{j,k_j}X_{i^{\prime},k_i^{\prime}}X_{l^{\prime},k_l^{\prime}}X_{i^{\prime},k_i^{\prime}})]\nonumber\\
 =& \frac{1}{(d+1)^3}\mathbf{E}[\Tr(X_{j,k_j}\rho_{i,k_i}X_{j,k_j}\rho_{m,k_m})\Tr(X_{j,k_j}X_{i^{\prime},k_i^{\prime}}\rho_{l^{\prime},k_l^{\prime}}X_{i^{\prime},k_i^{\prime}})]\nonumber\\
\leq & \frac{1}{(d+1)^3} \mathbf{E}[\Tr(X_{j,k_j}\rho_{m}) \Tr(X_{i^{\prime},k_i^{\prime}}\rho_{l,k_l})]\nonumber\\
\sim &\mathcal{O}\left(\frac{1}{(d+1)^5}\right).
\end{align}

\textit{case 1.3} {$k_j =k_j^{\prime} $, $k_m^{\prime} = k_i^{\prime}$ and $k_m = k_i$}.
\begin{align}
&\mathbf{E}
[\Tr(X_\1X_\2X_\1X_\2)\Tr(X_\1X_{i^{\prime},k_i^{\prime}} X_{l^{\prime},k_l^{\prime}}X_{i^{\prime},k_i^{\prime}})]\nonumber\\
=&\frac{1}{(d+1)}\mathbf{E}[\Tr(X_{j,k_j}X_{i,k_i})\Tr(X_{j,k_j}X_{i^{\prime},k_i^{\prime}})\Tr(\rho_{l,k_l}X_{i^{\prime},k_i^{\prime}})]\nonumber\\
\leq& \frac{1}{d+1}\mathbf{E}[\Tr(X_{j,k_j}X_{i,k_i})\Tr(\rho_{l}X_{i^{\prime},k_i^{\prime}})]\nonumber\\
\sim &\mathcal{O}\left(\frac{1}{(d+1)^3}\right).
\end{align}

\textit{case 1.4} $k_j = k_j^{\prime} = k_l^{\prime}$.
\begin{align}
&\mathbf{E}[\Tr(X_\1X_\2X_\3X_\4)\Tr(X_\1X_{i^{\prime},k_i^{\prime}}X_\1X_{m^{\prime},k_m^{\prime}})]\nonumber\\
=&\frac{1}{(d+1)^5}\mathbf{E}[\Tr(X_\1\rho_i\rho_l\rho_m)\Tr(X_\1\rho_iX_\1\rho_m)]\nonumber\\
\sim& \mathcal{O}\left(\frac{1}{(d+1)^6}\right).
\end{align}

\textit{case 1.5} $k_j =k_j^{\prime} $ and $k_m^{\prime} = k_i^{\prime}$. 
\begin{align}
   & \mathbf{E}[\Tr(X_\1X_\2X_\3X_\4)\Tr(X_\1X_{i^{\prime},k_i^{\prime}}X_{l^{\prime},k_l^{\prime}}X_{i^{\prime},k_i^{\prime}})]\nonumber\\
   =& \frac{1}{(d+1)^4}\mathbf{E} [\Tr(X_\1\rho_i\rho_l\rho_m)\Tr(X_\1X_{i^{\prime},k_i^{\prime}}\rho_\3 X_{i^{\prime},k_i^{\prime}})]\nonumber\\
   \sim & \mathcal{O}\left( \frac{1}{(d+1)^6}\right).
\end{align}

\textbf{case 2}
Two in $\{(i,k_i), (j,k_j), (l,k_l), (m,k_m)\}$ is equal to two in $\{(i^{\prime},k_i^{\prime}), (j^{\prime},k_j^{\prime}), (l^{\prime},k_l^{\prime}), (m^{\prime},k_m^{\prime})\}$. Without loss of generality, we assume $j = j^{\prime}$, $k_j =k_j^{\prime} $, $i = i^{\prime}$ and $k_i = k_i^{\prime}$. Now it can be divided into four cases.

\textit{case 2.1} $k_j =k_j^{\prime} $ and $k_i =k_i^{\prime} $
\begin{align}
&\mathbf{E}[\Tr(X_\1X_\2X_\3X_\4)\Tr(X_\1X_\2X_{l^{\prime},k_l^{\prime}}X_{m^{\prime},k_m^{\prime}})]\nonumber\\
=&\frac{1}{(d+1)^4}\mathbf{E}[\Tr(X_\1X_\2\rho_l\rho_m)\Tr(X_\1X_\2\rho_l\rho_m)]\nonumber\\
\sim &\mathcal{O}\left(\frac{1}{(d+1)^6}\right).
\end{align}

\textit{case 2.2} $k_j =k_j^{\prime} = k_l $ and $k_i =k_i^{\prime} $.
\begin{align}
&\mathbf{E}[\Tr(X_\1X_\2X_\1X_\4)\Tr(X_\1X_\2X_{l^{\prime},k_l^{\prime}}X_{m^{\prime},k_m^{\prime}})]\nonumber\\
=&\frac{1}{(d+1)^3}\mathbf{E}[\Tr(X_\1X_\2X_\1\rho_m)\Tr(X_\1X_\2\rho_l\rho_m)]\nonumber\\
= &\frac{1}{(d+1)^3}\mathbf{E}[\Tr(X_\1X_\2)\Tr(X_\1\rho_\4)\Tr(X_\1X_\2\rho_l\rho_m)]\nonumber\\
\leq & \frac{1}{(d+1)^3}\mathbf{E}[\Tr(X_\1X_\2)\Tr(X_\1X_\2\rho_l\rho_m)]\nonumber\\
\leq& \frac{1}{(d+1)^3}\sqrt{\mathbf{E}[|\Tr(X_\1X_\2)|^2]\mathbf{E}[|\Tr(X_\1X_\2\rho_l\rho_m)|^2]}\nonumber\\
\sim & \mathcal{O}\left(\frac{1}{(d+1)^5}\right).
\end{align}

\textit{case 2.3} $k_j =k_j^{\prime} = k_l $ and $k_i =k_i^{\prime} = k_m^{\prime}$.
\begin{align}
    &\mathbf{E}[\Tr(X_\1X_\2X_\1X_\4)\Tr(X_\1X_\2X_{l^{\prime},k_l^{\prime}} X_\2)]\nonumber\\
    =&\frac{1}{(d+1)^2}\mathbf{E}[\Tr(X_\1X_\2X_\1\rho_m)\Tr(X_\1X_\2\rho_l X_{i^{\prime},k_i^{\prime}})]\nonumber\\
    =&\frac{1}{(d+1)^2}\mathbf{E}[\Tr(X_\1X_\2)\Tr(X_\1\rho_m)\Tr(X_\1X_\2)\Tr(\rho_lX_\2)]\nonumber\\
    \leq&\frac{1}{(d+1)^2}\mathbf{E}|[\Tr(X_\1X_\2)|^2]\nonumber\\
    \sim & \mathcal{O}\left(\frac{1}{(d+1)^4}\right).
\end{align}

\textit{case 2.4} $k_i =k_i^{\prime} = k_m^{\prime} $ and $k_j =k_j^{\prime} $.
\begin{align}
    &\mathbf{E}[\Tr(X_\1X_\2X_\3X_\4)\Tr(X_\1X_\2X_{l^{\prime},k_l^{\prime}} X_\2)]\nonumber\\
    =&\frac{1}{(d+1)^3}\mathbf{E}[\Tr(X_\1X_\2\rho_l\rho_m)\Tr(X_\1X_\2\rho_lX_\2)]\nonumber\\
    \sim & \mathcal{O}\left(\frac{1}{(d+1)^5}\right).
\end{align}

\textit{case 2.5} $k_j =k_j^{\prime} = k_l $ and $k_i =k_i^{\prime} = k_m$.
\begin{align}
    &\mathbf{E}[\Tr(X_\1X_\2X_\1X_\2)\Tr(X_\1X_\2X_{l^{\prime},k_l^{\prime}} X_{m^{\prime},k_m^{\prime}})]\nonumber\\
    =&\frac{1}{(d+1)^2}\mathbf{E}[\Tr(X_\1X_\2X_\1X_\2)\Tr(X_\1X_\2\rho_l \rho_m)]\nonumber\\
    \leq&\frac{1}{(d+1)^2}\mathbf{E}[|\Tr(X_\1X_\2)|^2]\nonumber\\
    \sim & \mathcal{O}\left(\frac{1}{(d+1)^4}\right).
\end{align}

\textbf{case 3}
Two in $\{(i,k_i), (j,k_j), (l,k_l), (m,k_m)\}$ is equal to two in $\{(i^{\prime},k_i^{\prime}), (j^{\prime},k_j^{\prime}), (l^{\prime},k_l^{\prime}), (m^{\prime},k_m^{\prime})\}$. Without loss of generality, we assume $j = j^{\prime}$, $k_j =k_j^{\prime} $, $l = l^{\prime}$ and $k_l = k_l^{\prime}$. Now it can be divided into four cases.

\textit{case 3.1} $k_j = k_j^{\prime}$ and $k_l = k_l^{\prime}$
\begin{align}
    &\mathbf{E}[\Tr(X_\1X_{i, k_i} X_\3X_{m,k_m} )\Tr(X_\1X_{i^{\prime}, k_i^{\prime}} X_\3X_{m^{\prime}, k_m^{\prime}} )]\nonumber\\
    \sim & \mathcal{O}\left(\frac{1}{(d+1)^6}\right).
\end{align}

\textit{case 3.2} $k_j = k_j^{\prime}$, $k_l = k_l^{\prime}$ and $k_i^{\prime} = k_m^{\prime}$
\begin{align}
    &\mathbf{E}[\Tr(X_\1X_{i, k_i} X_\3X_{m, k_m} )\Tr(X_\1X_{i^{\prime}, k_i^{\prime}}X_\3X_{i^{\prime}, k_i^{\prime}})] \nonumber\\
    =&\frac{1}{(d+1)^2}\mathbf{E}[\Tr(X_\1\rho_iX_\3\rho_m)Tr(X_\1X_{i^{\prime}, k_i^{\prime}})Tr(X_\3X_{i^{\prime}, k_i^{\prime}})]\nonumber\\
    \leq & \frac{1}{(d+1)^3}\sqrt{\mathbf{E}[|\Tr(X_\1X_{i^{\prime}, k_i^{\prime}})|^2]}\sqrt{\mathbf{E}[|\Tr(X_\3X_{i^{\prime}, k_i^{\prime}})|^2]}\nonumber\\
    \sim &\mathcal{O}\left(\frac{1}{(d+1)^5}\right).
\end{align}

\textit{case 3.3} $k_j = k_j^{\prime}$, $k_l = k_l^{\prime}$, $k_i = k_m$ and $k_i^{\prime} = k_m^{\prime}$.
\begin{align}
&\mathbf{E}[\Tr(X_\1X_\2X_\3X_\2)
\Tr(X_\1X_{i^{\prime}, k_i^{\prime}}X_\3X_{i^{\prime}, k_i^{\prime}})\nonumber\\
=&\mathbf{E} [\Tr(X_\1X_\2)\Tr(X_\2X_\3)\Tr(X_\1X_{i^{\prime}, k_i^{\prime}})\Tr(X_{i^{\prime}, k_i^{\prime}}X_\3)]\nonumber\\
    \sim &\mathcal{O}\left(\frac{1}{(d+1)^4}\right).
\end{align}

\textbf{case 4}
Three in $\{(i,k_i), (j,k_j), (l,k_l), (m,k_m)\}$ is equal to three in $\{(i^{\prime},k_i^{\prime}), (j^{\prime},k_j^{\prime}), (l^{\prime},k_l^{\prime}), (m^{\prime},k_m^{\prime})\}$. Without loss of generality, we assume $j = j^{\prime}$, $k_j =k_j^{\prime} $, $l = l^{\prime}$, $k_l = k_l^{\prime}$, $i = i^{\prime}$ and $k_i = k_i^{\prime}$ Now it can be divided into four cases.

\textit{case 4.1} $k_j = k_j^{\prime}$, $k_i = k_i^{\prime}$ and $k_l = k_l^{\prime}$.
\begin{align}
&\mathbf{E}\Tr(X_\1X_\2X_\3X_\4)\Tr(X_\1X_\2X_\3X_{m^{\prime},k_m^{\prime}})\nonumber\\
=&\frac{1}{(d+1)^2}\mathbf{E}[\Tr(X_\1X_\2X_\3\rho_m)\Tr(X_\1X_\2X_\3\rho_m)]\nonumber\\
\leq &\frac{1}{(d+1)^3} \sqrt{\mathbf{E}[|\Tr(X_\1X_\2)|^2}]\sqrt{\mathbf{E}[|\Tr(X_\2X_\3)|^2}]\nonumber\\
    \sim &\mathcal{O}\left(\frac{1}{(d+1)^5}\right).
\end{align}

\textit{case 4.2} $k_j = k_l = k_j^{\prime} = k_l^{\prime}$ and $k_i = k_i^{\prime}$.
\begin{align}
&\mathbf{E}[\Tr(X_\1X_\2X_\1X_\4)\Tr(X_\1X_\2X_\1X_{m^{\prime},k_m^{\prime}})]\nonumber\\
=&\frac{1}{(d+1)^2}\mathbf{E}[\Tr(X_\1X_\2X_\1\rho_m)\Tr(X_\1X_\2X_\1\rho_m)]\nonumber\\
\leq &\frac{1}{(d+1)^2}\mathbf{E}[|\Tr(X_\1X_\2)|^2]\nonumber\\
    \sim &\mathcal{O}\left(\frac{1}{(d+1)^4}\right).
\end{align}

\textit{case 4.3} $k_j = k_l = k_j^{\prime} = k_l^{\prime}$ and $k_i = k_i^{\prime} = k_m$.
\begin{align}
    &\mathbf{E}[\Tr(X_\1X_\2X_\1X_\2)\Tr(X_\1X_\2X_\1X_{m^{\prime},k_m^{\prime}})]\nonumber\\
    =&\frac{1}{(d+1)}\mathbf{E}[\Tr(X_\1X_\2X_\1X_\2)\Tr(X_\1X_\2X_\1\rho_m)]\nonumber\\
    \leq & \frac{1}{(d+1)}\mathbf{E}[|\Tr(X_\1X_\2)|^2]\nonumber\\
        \sim &\mathcal{O}\left(\frac{1}{(d+1)^3}\right).
\end{align}

\textit{case 4.4} $k_j = k_j^{\prime}$, $k_l = k_l^{\prime}$ and $k_i = k_i^{\prime} = k_m$.
\begin{align}
    &\mathbf{E}[\Tr(X_\1X_\2X_\3X_\2)\Tr(X_\1X_\2X_\3X_{m^{\prime},k_m^{\prime}})]\nonumber\\
    =&\frac{1}{(d+1)}\mathbf{E}[\Tr(X_\1X_\2X_\3X_\2)\Tr(X_\1X_\2X_\3\rho_m)]\nonumber\\
    \leq & \frac{1}{(d+1)^2}\mathbf{E}[|\Tr(X_\1X_\2)|^2|\Tr(X_\2X_\3)|^2]\nonumber\\
        \sim &\mathcal{O}\left(\frac{1}{(d+1)^4}\right).
\end{align}

\textbf{case 5}
Four pairs $\{(i,k_i), (j,k_j), (l,k_l), (m,k_m)\}$ are equal $\{(i^{\prime},k_i^{\prime}), (j^{\prime},k_j^{\prime}), (l^{\prime},k_l^{\prime}), (m^{\prime},k_m^{\prime})\}$, respectively. Now it can be divided into three cases.
\textit{case 5.1} $k_j = k_j^{\prime}$, $k_i = k_i^{\prime}$, $k_l= k_l^{\prime}$ and $k_m = k_m^{\prime}$.
\begin{align}
    &\mathbf{E}[\Tr(X_\1X_\2X_\3X_\4)\Tr(X_\1X_\2X_\3X_\4)]\nonumber\\
    =&\mathbf{E}[\Tr(X_\1X_\2)\Tr(X_\2X_\3)\Tr(X_\3X_\4)\Tr(X_\4X_\1)]\nonumber\\
        \sim &\mathcal{O}\left(\frac{1}{(d+1)^4}\right).
\end{align}

\textit{case 5.2} $k_j = k_j^{\prime} = k_l = k_l^{\prime}$, $k_i = k_i^{\prime}$ and $k_m = k_m^{\prime}$.
\begin{align}
    &\mathbf{E}[\Tr(X_\1X_\2X_\1X_\4)\Tr(X_\1X_\2X_\1X_\4)]\nonumber\\
    \leq&\frac{1}{d+1}\mathbf{E}[\Tr(X_\1X_\4)\Tr(X_\4X_\1)]\nonumber\\
        \sim &\mathcal{O}\left(\frac{1}{(d+1)^3}\right).
\end{align}

\textit{case 5.3} $k_j = k_j^{\prime} = k_l = k_l^{\prime}$ and $k_i = k_i^{\prime} = k_m = k_m^{\prime}$.
\begin{align}
    &\mathbf{E}[\Tr(X_\1X_\2X_\1X_\2)\Tr(X_\1X_\2X_\1X_\2)]\nonumber\\
    =&\mathbf{E}|[\Tr(X_\1X_\2)|^4]\nonumber\\
    \leq&\mathbf{E}[|\Tr(X_\1X_\2)|^2]\nonumber\\
            \sim &\mathcal{O}\left(\frac{1}{(d+1)^2}\right).
\end{align}

Next, we turn to calculate the variance of the first term of Eq.~\ref{eq: div bound}. The variance can be expanded as
\begin{align}
    &Var[\langle\langle\Delta (\rho_i\rho_j)|H\otimes I_d|\Delta(\rho_l\rho_m)\rangle\rangle ] \nonumber\\ & \leq \lambda_{max}^2  Var[\langle\langle\Delta (\rho_i\rho_j)|\Delta(\rho_l\rho_m\rangle\rangle ])\nonumber\\
    &\leq \lambda_{max}^2 \mathbf{E} |\langle\langle\Delta (\rho_i\rho_j)|\Delta(\rho_l\rho_m)\rangle\rangle|^2\nonumber\\
    & =  \lambda_{max}^2( \mathbf{E} |\langle\langle\Tilde{\rho}_i\Tilde{\rho}_j|\Tilde{\rho}_l\Tilde{\rho}_m\rangle\rangle|^2 - |\mathbf{E} (\langle\langle\Tilde{\rho}_i\Tilde{\rho}_j|\Tilde{\rho}_l\Tilde{\rho}_m\rangle\rangle)|^2,
    \label{eq: var}
\end{align}
where the first term of Eq.~\ref{eq: var} can be expanded as

    \begin{align}
     \mathbf{E} |\langle\langle\Tilde{\rho}_i\Tilde{\rho}_j|\Tilde{\rho}_l\Tilde{\rho}_m\rangle\rangle|^2 = \frac{1}{M^8}\sum_{k_1,k_2,k_3,k_4,k_1^{\prime},k_2^{\prime},k_3^{\prime},k_4^{\prime}}\mathbf{E}\langle\langle\rho_{1,k_1}\rho_{2,k_2}|\rho_{3,k_3}\rho_{4,k_4}\rangle\rangle\langle\langle\rho_{1,k_1^{\prime}}\rho_{2,k_2^{\prime}}|\rho_{3,k_3^{\prime}}\rho_{4,k_4^{\prime}}\rangle\rangle.
\end{align}

Similar to when calculating the deviation, here we only need to consider the situation where at least one element in $\{(i,k_i), (j,k_j), (l,k_l), (m,k_m)\}$  is equal to some element in $\{(i^{\prime},k_i^{\prime}), (j^{\prime},k_j^{\prime}), (l^{\prime},k_l^{\prime}), (m^{\prime},k_m^{\prime})\}$. Because other situations will be canceled out with the second term of Eq,~\ref{eq: var}.

Further expansion can be obtained

\begin{align}
&\mathbf{E}\langle\langle\rho_{1,k_1}\rho_{2,k_2}|\rho_{3,k_3}\rho_{4,k_4}\rangle\rangle\langle\langle\rho_{1,k_1^{\prime}}\rho_{2,k_2^{\prime}}|\rho_{3,k_3^{\prime}}\rho_{4,k_4^{\prime}}\rangle\rangle\nonumber\\
= & (d+1)^8\mathbf{E}\Tr(X_\1X_\2X_\3X_\4)\Tr(X_{j^{\prime},k_j^{\prime}}X_{i^{\prime},k_i^{\prime}}X_{l^{\prime},k_l^{\prime}}
X_{m^{\prime},k_m^{\prime}})\nonumber\\
- &(d+1)^7\mathbf{E}\Tr(X_\2X_\3X_\4)\Tr(X_{j^{\prime},k_j^{\prime}}X_{i^{\prime},k_i^{\prime}}X_{l^{\prime},k_l^{\prime}}
X_{m^{\prime},k_m^{\prime}})\dots\nonumber\\
+ & 
(d+1)^6\mathbf{E}\Tr(X_\3X_\4)\Tr(X_{j^{\prime},k_j^{\prime}}X_{i^{\prime},k_i^{\prime}}X_{l^{\prime},k_l^{\prime}}
X_{m^{\prime},k_m^{\prime}})\dots\nonumber\\
+&\dots\nonumber\\
=& \sum_{p=0}^{p=8}(d+1)^p F_p,
\label{eq: expand}
\end{align}
where $F_p$ packages all items starting with $(d+1)^p$, that is the combination of $p$ different $X$s.

According to Lemma~\ref{lemma1}, the expectation $F_p$ is determined by its degree of freedom $f$, i.e., $F_p\sim \mathcal{O}(1/d^f)$. When $p$ decreases by $1$, its degree of freedom $f$ may decrease by $1$ or remain unchanged. Therefore, the largest term in Eq.~\ref{eq: expand} is the first term $(d+1)^8 F_p \sim\mathcal{O}((d+1)^{8-f})$. After traversing $M^8$ combinations, there are $M^f$ combinations that satisfy the degree of freedom $f$, so the final variance is $\sim \mathcal{O}(d^{8-f}/M^{8-f})$. 

\end{document}